\documentclass[prc,twocolumn,showpacs,preprintnumbers,amsmath,amssymb]{revtex4}
\usepackage{epsfig}
\usepackage{graphicx}
\usepackage{xcolor}
\usepackage{ulem}
\begin{document}

\title{Fermi motion effects in electroproduction of hypernuclei}
\author{P. Byd\v{z}ovsk\'y$^a$, D. Denisova$^{a,b}$, D. Skoupil$^a$, 
and P. Vesel\'y$^a$}
\affiliation{$^a$Nuclear Physics Institute, ASCR, \v{R}e\v{z}/Prague, 
Czech Republic\\
$^b$Institute of Particle and Nuclear Physics, Faculty of Mathematics 
and Physics, Charles University, Prague, Czech Republic}

\date{\today }

\begin{abstract}
In a previous analysis of electroproduction of hypernuclei the cross sections 
were calculated using the distorted-wave impulse approximation where the momentum 
of the initial proton in the nucleus was set to zero (the ``frozen-proton'' 
approximation). In this paper we go beyond this approximation assuming a non zero 
effective proton momentum due to proton Fermi motion inside the target nucleus 
discussing also other kinematical effects. To this end we have derived a more 
general form of the two-component elementary electroproduction amplitude 
(Chew-Goldberger-Low-Nambu like) which allows its use in a general reference frame 
moving with respect to the nucleus rest frame. The effects of Fermi motion were 
found to depend on kinematics and elementary amplitudes. The largest effects were 
observed in the contributions from the longitudinal and interference parts of the 
cross sections. The extension of the calculations beyond the frozen-proton 
approximation improved the agreement of predicted theoretical cross sections with 
experimental data, and, once we assumed the optimum on-shell approximation, we were 
able to remove an inconsistency which was previously present in the calculations.
\end{abstract}

\maketitle
\section{Introduction}
Studying the production of $\Lambda$ hypernuclei provides important information 
on details of the $\Lambda$N interaction, particularly on its spin-dependent part 
that is difficult to investigate in free $\Lambda$N scattering~\cite{rijken99}. 
In fact, the effective $\Lambda$N interaction can be determined from hypernuclear 
spectra obtained from various reactions induced by hadron (mainly $\pi^+$ and 
$K^-$) and electron beams~\cite{GHM2016}. Moreover, precise measurements of 
the production cross sections provide information on the hypernuclear production 
mechanism and the dynamics of the elementary production reaction. Whereas 
$\gamma$-ray hypernuclear spectra are measured with very high precision (a few 
keV),  the spectra of (e, e$^\prime$ K$^+$) reactions from nuclei are 
obtained only with resolution of several hundreds of keV, which is better than 
in the hadron-induced reactions~\cite{hashtam06} but which still makes studying 
the fine splitting within the multiplets problematic. On the other hand, the 
reaction spectroscopy allows one to study also higher energy excited states, 
{\it e.g.}, even between the nucleon emission threshold and the $\Lambda$ 
emission threshold which is not possible in $\gamma$-ray spectroscopy.  

The electroproduction of strangeness is characterized by a large three-momentum 
transfer to the $\Lambda$ ($\approx$250 MeV/c), large angular momentum transfer 
$J$, and strong spin-flip terms, even at almost zero production 
angles~\cite{hashtam06}. The latter results in the dominance of the highest-spin 
states in the multiplets. Moreover, the kaon production occurs on a proton 
in contrast to a neutron in the (K$^-$, $\pi^-$) and ($\pi^+$, K$^+$) reactions 
allowing one to study different hypernuclei and charge-dependent effects from 
a comparison of mirror hypernuclei (Charge Symmetry Breaking). An important merit 
of electroproduction in view of its theoretical description is that the 
electro-magnetic part of the interaction is well known. A systematic experimental 
study of high-resolution hypernuclear spectroscopy in (e, e$^\prime$ K$^+$) has 
been performed in Halls A~\cite{archival} and C~\cite{HallC} at Jefferson 
Laboratory (JLab). However, a reliable analysis of these data requires good 
understanding of the dynamics of the process and uncertainties arising from 
various approximations in a model.   

Theoretical calculations of the cross sections in photo- and electroproduction 
of hypernuclei~\cite{Cotanch,Bennhold,SF1994,HLee,Lenske,PTP2010,NPA2012,
Motoba,AIP2018} have been performed in the distorted-wave impulse 
approximation (DWIA) mainly assuming that the initial proton is at rest 
with respect to the nucleus. This ``frozen-proton'' approximation made it 
possible to use the nonrelativistic two-component Chew-Goldberger-Low-Nambu 
(CGLN) form of the elementary-production amplitude in the proton laboratory 
frame~\cite{SF1994}, which significantly simplifies the calculations. 
However, as the proton is moving inside of the nucleus with a momentum about 
150~MeV/c (in $^{12}$C), which is comparable to a momentum transfer of about 
250~MeV/c, it is advisable to estimate effects arising from this proton Fermi 
motion. Note that to this end, in the approaches with a non relativistic 
shell-model description of the nuclear structure, one needs to derive the 
two-component form of the elementary electroproduction amplitude for a moving 
proton, {\it i.e.}, the CGLN-like form in a general Lorentz frame moving with 
respect to the target-nucleus center of gravity. 

The Fermi motion was included, \textit{e.g.}, in the elastic scattering of 
protons~\cite{FM-proton} and pions~\cite{FM-pion} from nuclei, via 
full folding (Fermi averaging) in the first-order optical potential. 
Note that here one needs an off-energy-shell extension of the elementary 
amplitude which introduces additional uncertainty in the calculation. 
Fermi motion effects were found to be important in calculating the cross 
sections in pion photoproduction off $^{10}$B~\cite{Sato}.
On the other hand, in the DWIA analysis of the (K$^-$, $\pi^-$) reaction 
on $^{12}$C~\cite{Zofka} it was shown that accounting for the Fermi 
motion results only in a few percent reduction of the cross sections for 
hypernuclear production.  
The effects from motion of the target nucleon were also included  
in the DWIA calculations of the cross section in the ($\pi^+$, K$^+$) and 
(K$^-$, $\pi^-$) reactions on the $^{12}$C target assuming optimal 
Fermi averaging of the on-shell elementary amplitude~\cite{Harada}.
This averaging of the amplitude leads to improvement in the data description.
A Fermi averaged amplitude was also used in the study of formation 
of p-shell hypernuclei in the (K$^-$, $\pi^-$) reactions~\cite{Auerbach}.
Nonlocalities in photoproduction of hypernuclei arising from Fermi motion 
were included in Refs.~\cite{Bennhold,Lenske} using relativistic 
nuclear models. These effects were found to make changes of about 20\% 
(and more) in the cross sections for the high-spin transitions which 
dominate in a multiplet~\cite{Bennhold}.

In this paper we study effects of Fermi motion of the initial proton 
in DWIA calculations of the cross sections in 
$^{12}$C(e,e$^\prime$K$^+$)$^{12}_{~\Lambda}$B. 
In order to avoid uncertainties related to an off-energy-shell extension of 
the elementary amplitude, constructed for production on a free 
proton~\cite{SB18,SLA}, the amplitude is considered on-shell in the optimal 
factorisation approximation. This approach was already used in our previous 
calculations~\cite{NPA2012,archival,AIP2018} and here we just go beyond 
the frozen-proton approximation. We also suggest a solution with an  
``optimum'' proton momentum which allows us to use the on-shell 
amplitude fulfilling simultaneously energy conservation in 
the many-body system. 
The Fermi-motion effects will be demonstrated on the angle and energy 
dependent cross sections and the new results will be also compared with 
available experimental data and our previous results from Ref.~\cite{archival}.

Including the elementary amplitude for a non zero proton momentum required 
derivation of the two-component form of the amplitude in a general reference 
frame. This two-component, CGLN-like, form of the covariant amplitude  
derived in the field-theoretical framework~\cite{SB16,SB18} is necessary 
because the nuclear and hypernuclear structure is described in the 
nonrelativistic quantum-mechanical frame, {\it e.g.}, in the shell model.
To our knowledge, such a general two-component amplitude for 
electroproduction of pseudoscalar mesons is not available in 
the literature and therefore we have derived the formulas by ourselves.

The paper is organised as follows: in the next section, we briefly 
describe the formalism of DWIA and the two-component elementary 
amplitude in a general reference frame. Results showing the Fermi 
motion effects in the cross sections are discussed in Sec. III. 
In this section we also provide updated theoretical predictions for 
the $^{12}$C, $^9$Be, and $^{16}$O targets in comparison with the data 
and previous results published in Ref.~\cite{archival}. 
A summary and conclusions are given in Sec. IV. More details on the 
formalism, formulas and derivations, are given in Appendices A, B, and C.    
%
%
\section{Formalism}
In this section we provide a basic formalism of the DWIA suitable for description of 
electroproduction of hypernuclei.  
%
\subsection{Optimal factorization approximation}
\begin{figure}[htb]
\begin{center}
\includegraphics[width=0.18\textwidth,angle=270]{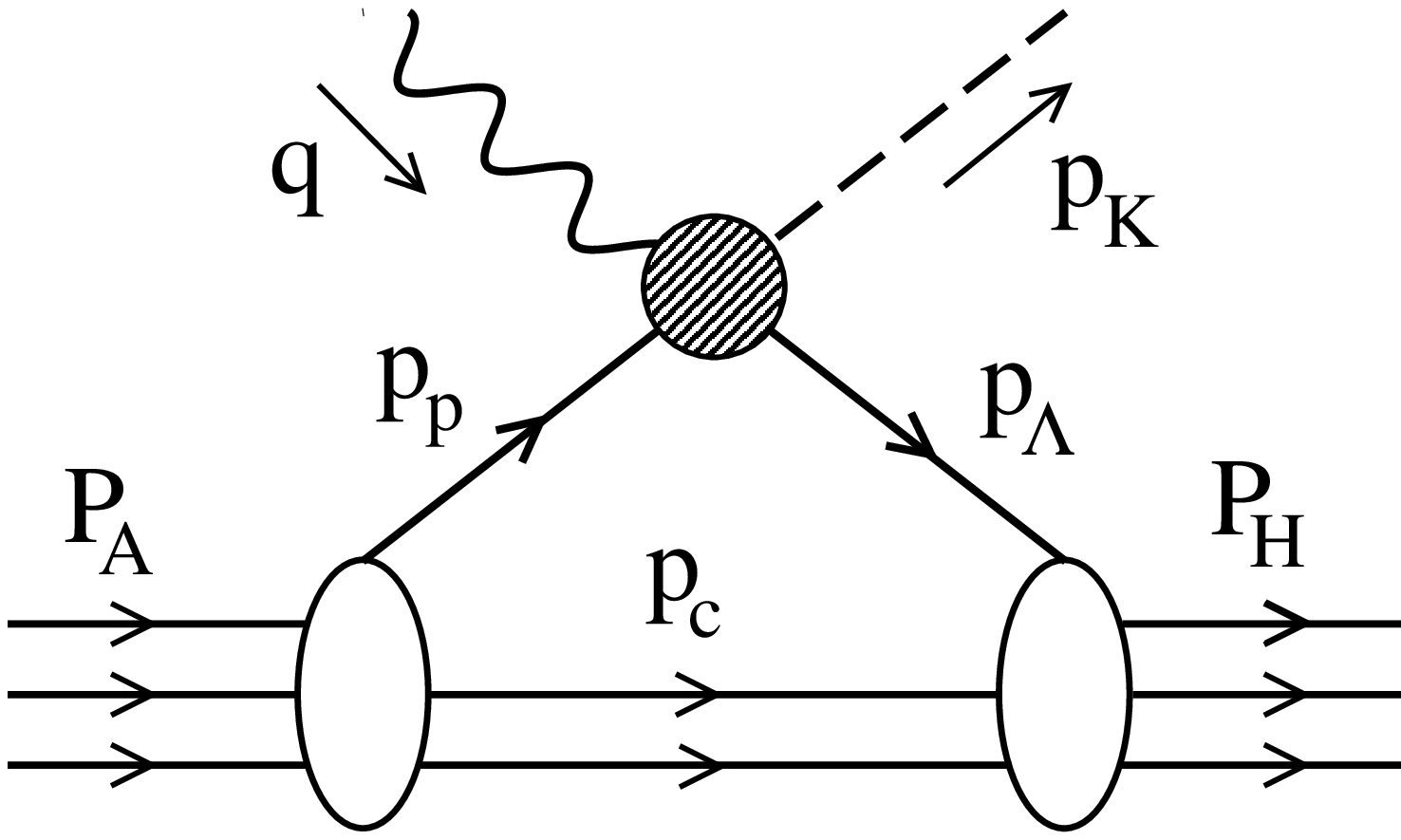}
\end{center}
\caption{A schematic representation of the amplitude for photoproduction 
 of a hypernucleus (H) induced by virtual photons in the plane-wave impulse 
 approximation. The energy-momenta of the nucleus, denoted by A, and of the 
 hypernucleus, denoted by H, are marked with capital letters 
 ${\rm P}=(E,\vec{P}\,)$ and those of intermediate systems with lowercase letters.}
\label{impulse-approximation}
\end{figure}

Production of hypernuclei by a virtual photon associated with a kaon 
in the final state, 
\begin{equation}
\gamma_{\sf v}(\rm q) + {\rm A}(P_{\rm A})\longrightarrow 
{\rm H}(P_{\rm H})+{\rm K^+}(p_{\rm K})\,,
\label{electro}
\end{equation}
where the corresponding four-momenta are given in the parentheses,  
can be satisfactorily described in the impulse approximation (IA) where  
the elementary reaction takes place on individual protons bound in 
the nucleus as shown in Fig.~\ref{impulse-approximation}. 
%
%
This approach is justified because the photon and kaon momenta are supposed 
to be rather high ($\approx 1~\text{GeV}$), {\it e.g.}, in the JLab 
experiments~\cite{archival,HallC}, and this clearly separates the elementary 
production (the 2-body process) from the many-body reaction (\ref{electro}). 
The cross section for production of the ground and excited states of a hypernucleus 
depends on the many-particle matrix element between the nonrelativistic wave 
functions of the target nucleus ($\Psi_A$) and the hypernucleus ($\Psi_H$)
\begin{equation}
{\sf M}_\mu= \langle \Psi_H | \langle \chi_K | 
\sum_{j=1}^{\sf Z}\;\hat{\rm \!J}_\mu(j)\,|\chi_\gamma\rangle|\Psi_A\rangle\,,
\label{matrix_element}
\end{equation}
where $\chi_K$ is taken as a kaon plane (PWIA) or distorted (DWIA) 
wave function and $\hat{\rm \!J}_\mu(j)$ is an elementary operator 
for $\Lambda$ production on the $j$th proton. 
In the one-photon approximation without Coulomb distortion of 
initial and final electrons, the virtual-photon wave function 
$\chi_\gamma$ is taken as the plane wave.  
Due to the symmetry of the nuclear wave function, we can replace 
the sum over all protons with ${\sf Z}\;\hat{\rm \!J}_\mu$ where 
${\sf Z}$ is the atomic number of the target nucleus. 
Denoting the intermediate momenta of the 
proton and $\Lambda$ as $\vec{p}_p$ and 
$\vec{p}_\Lambda$, respectively, the momentum transfer as  
$\vec{\Delta}=\vec{q}-\vec{p}_K$, and considering 
translational invariance, we introduce the elementary amplitude 
${\cal J}_\mu (\vec{p}_K, \vec{q} ,\vec{p}_p )$
\begin{equation}
\langle \vec{p}_K,\vec{p}_\Lambda\,|\,\hat{\rm \!J}_\mu|\,
\vec{q},\vec{p}_p\rangle = (2\pi)^3\,
\delta^{(3)}(\vec{p}_\Lambda-\vec{p}_p-\vec{\Delta})\,
{\cal J}_\mu ,
\label{elamp}
\end{equation}
which has to be expressed in the two-component form to match it with 
the nonrelativistic nuclear and hypernuclear wave functions 
in Eq.~(\ref{matrix_element}). Similarly, we introduce the hypernuclear  
production amplitude ${\cal T}_\mu (\vec{p}_K, \vec{q} ,\vec{P}_A )$
\begin{equation}
{\sf M}_\mu= (2\pi)^3\,
\delta^{(3)}(\vec{P}_H -\vec{P}_A-\vec{\Delta})\,
{\cal T}_\mu \,.
\label{prodamp}
\end{equation}
In the nucleus-rest (lab) frame, this amplitude is
\begin{eqnarray}
\!\!\!{\cal T_\mu}\!&=&\!\!{\sf Z}\!\int\!d^3\xi_\Lambda\,d^3\xi_p\,
\exp\,(iB\,\vec{\Delta}\cdot\vec{\xi}_\Lambda)\;
\chi^*_K(\vec{p}_{KH},B\vec{\xi}_\Lambda)\nonumber\\
&\times&\int\!\frac{d^3p_p}{(2\pi)^3}\exp\biggl[i\vec{p}_p\cdot 
(\vec{\xi}_\Lambda-\vec{\xi}_p)\biggr]
\,{\sf Tr}\!\left[\!\frac{}{}\!{\cal J}_\mu (\vec{p}_K, \vec{q} ,\vec{p}_p )
\right. \nonumber\\
&\times&\int\!d^3\xi_1\,...\, d^3\,\xi_{A-2}\;
\Phi^*_{\sf H}(\vec{\xi}_1,\,...\,\vec{\xi}_{A-2},\vec{\xi}_\Lambda) 
\nonumber\\
&\times&\;\;\;\left.\Phi_{\sf A}(\vec{\xi}_1,\,...\,\vec{\xi}_{A-2},\vec{\xi}_p)
\right]\,.
\label{matrix-4}
\end{eqnarray}
where $B=(A-1)/(A-1+m_\Lambda/m_p)$, $\xi$'s are the Jacobi coordinates, and 
$\chi^*_K$ describes the kaon distortion ($\chi^*_K=1$ in PWIA) that depends 
on the kaon relative momentum with respect to the hypernucleus, 
$\vec{p}_{KH}$. The trace is over the proton and $\Lambda$ spin as 
${\cal J}_\mu$ is the $2\times 2$ matrix in that space. 
The integral over the proton momentum $\vec{p}_p$ includes averaging over 
the Fermi motion of the target protons inside the nucleus. 
However, we can assume that a dependence on $\vec{p}_p$ in the amplitude 
${\cal J}_\mu$ is much smoother than that in the phase factor weighted 
by the nuclear wave functions and take the elementary amplitude out 
of the integral for some effective (optimal) value of $\vec{p}_p$;  
{\it i.e.}, we can replace  
${\cal J}_\mu (\vec{p}_K, \vec{q} ,\vec{p}_p)$ with  
${\cal J}_\mu (\vec{p}_K, \vec{q} ,\vec{p}_{\sf eff})$.  
The integration over $\vec{p}_p$ then gives a $\delta$-function which 
allows integration over $\vec{\xi}_\Lambda$. Finally we arrive at 
an expression for the lab amplitude in the optimal factorization 
approximation (OFA) (see also~\cite{FM-proton})  
\begin{eqnarray}
&&\!\!\!\!\!\!{\cal T_\mu} = 
{\sf Z}\,{\sf Tr}\left[ {\cal J}_\mu 
(\vec{p}_K,\vec{q},\vec{p}_{\sf eff})
\!\int\!d^3\xi\,e^{(iB\,\vec{\Delta}\cdot\vec{\xi})}
\,\chi^*_K(\vec{p}_{KH},B\vec{\xi})\right.
\nonumber\\
&\times&\!\!\!\!\left.\int\!d^3\xi_1... d^3\,\xi_{A-2}
\Phi^*_{\sf H}(\vec{\xi}_1,...\,\vec{\xi}_{A-2},\vec{\xi}\,)
\Phi_{\sf A}(\vec{\xi}_1,...\,\vec{\xi}_{A-2},\vec{\xi}\,)\right],
\nonumber\\
\label{matrix-5}
\end{eqnarray} 
where we omit a subscript at the integration variable $\vec{\xi}$ 
(as $\vec{\xi}_p=\vec{\xi}_\Lambda$) and $\vec{p}_{\sf eff}$ is an effective 
proton momentum. Recall that $\vec{\xi}$ is the relative particle-core 
coordinate. 

In the previous calculations, see {\it e.g.} Refs.~\cite{SF1994,archival,Motoba}, 
the effective momentum was set to zero in the so-called frozen-proton 
approximation. This choice provides us with a simple expression for 
the two-component (CGLN) elementary amplitude in the proton laboratory frame, 
see Eq.~$(4.3)$ 
in \cite{SF1994}. Here the CGLN amplitude consists only of six structures. 
In the more general case with a non zero $\vec{p}_{\sf eff}$ one needs 
a more general CGLN-like form. In the next section we are going to derive 
this form which will allow us to go beyond the frozen-proton approximation.
%
\subsection{Two-component elementary amplitude}
\label{newCGNL-amplitude}
Here we will present the two-component formalism for kaon electroproduction 
but it can be also applied for electroproduction of any pseudo-scalar meson.

The invariant amplitude  of K$^+$ production on 
a free proton induced by a virtual photon
\begin{equation}
\gamma_{\sf v}(q,\varepsilon) + {\sf p}(p_p,\eta_p)\longrightarrow 
\Lambda(p_\Lambda,\eta_\Lambda)+{\sf K^+}(p_K)\,,
\label{elementary}
\end{equation} 
can be expressed via six gauge invariant (GI) operators   
%
\begin{equation}
{\cal M}^\mu \varepsilon_\mu= \bar{u}(p_\Lambda ,\eta_\Lambda )\, 
\gamma_5\, \sum_{j=1}^6 
{\sf M}_j\,A_j(q^2,s,t)\,u(p_p,\eta_p)
\label{invariant}
\end{equation} 
where, in the one-photon approximation, the electron part enters 
via the four-vector 
$\varepsilon_\mu = \frac{e}{q^2}\bar{u}(p^\prime_e)\,\gamma_\mu u(p_e)$  
and the mass of the virtual photon is $q^2<0$. 
Dirac bispinors $\bar{u}_\Lambda$ and $u_p$ with spin projections 
$\eta_\Lambda$ and $\eta_p$ are  for the $\Lambda$ and proton, respectively. 
The scalar amplitudes $A_j$ are functions of Mandelstam variables 
$s=(q +p_p)^2$ and $t=(q-p_K)^2$ and describe the reaction dynamics. 
They are obtained from a decomposition of contributions from Feynman  
diagrams~\cite{SB16,SB18}. 
The GI operators ${\sf M}_j$, composed of $q$, $p_p$, $p_\Lambda$, $\varepsilon_\mu$, 
and $\gamma$-matrixes, assure that the invariant amplitude (\ref{invariant}) 
fulfills the Ward identity ${\cal M}\cdot q = 0$. 
Formulas for ${\sf M}_j$ are given in Eqs.~(17) of Ref.~\cite{SB16}. 
Specific expressions for the scalar amplitudes $A_j(s,t)$ and more 
details on the formalism in photo- and electroproduction of kaons 
can be found in Refs.~\cite{SB16,SB18}.

The invariant amplitude (\ref{invariant}) can be used to calculate  
observable quantities of the process (\ref{elementary}) 
in any reference frame but it cannot be directly used 
to calculate the many-particle matrix element (\ref{matrix_element}). 
For computing this matrix element we need a one-body transition operator 
on the nonrelativistic proton-hyperon Hilbert space written in the 
two-component formalism, {\it i.e.} using Pauli matrices. 
Only then can we fold the operator with the nuclear and hypernuclear  
nonrelativistic wave functions that are expressed by Pauli spinors. 
Moreover, as we see in Eq.~(\ref{matrix-5}) we need the elementary 
amplitude, in general, for a nonzero proton momentum 
$\vec{p}_{\sf eff}$. Formulas for this two-component (CGLN) amplitude 
in special cases are already available in literature, see 
{\it e.g.} Ref.~\cite{SF1994} for either the laboratory 
($\vec{p}_\text{eff}=0$) or the center-of-mass ($\vec{p}_\text{eff}=-\vec{q}$) 
frame. Here we provide the two-component form of the amplitude for arbitrary 
value of the proton momentum, {\it i.e.}, in an arbitrary frame. 

Due to the GI of the amplitude (\ref{invariant}) one can 
change the polarization vector  
$\epsilon_\mu=\varepsilon_\mu- \varepsilon_0\, q_\mu/q_0=
(\,0,\,\vec{\epsilon}\;)$, which sets the time component to zero,
and the amplitude can be written via a two-component amplitude 
$\vec{J}\cdot\vec{\epsilon}$,  
\begin{equation}
{\cal M}\cdot\varepsilon =  \bar{u}_\Lambda \gamma_5
\sum_{j=1}^6{\sf M}_j\,A_j \,u_p =
{\sf X}_\Lambda^\dag\ (\vec{J}\cdot\vec{\epsilon}\,)\ \;{\sf X}_p\,,
\label{reduction}
\end{equation}
where ${\sf X_\Lambda}^\dag$ and ${\sf X}_p$ are Pauli spinors. 
The new form $\vec{J}$  includes nonrelativistic structures composed from 
Pauli matrices and three-momenta, for example $\vec{q}$, $\vec{p}_K$, and 
$\vec{p}_p$. After some manipulations we can arrive at the amplitude in 
the two-component form for a nonzero proton momentum $\vec{p}_p$, 
\begin{eqnarray}
\vec{J}\cdot\vec{\epsilon} &=& G_1\,(\vec{\sigma}\cdot\vec{\epsilon}\,) +
G_2\,i(\vec{p}_p\times\vec{q}\cdot\vec{\epsilon}\,) +
G_3\,i(\vec{p}_K\times\vec{q}\cdot\vec{\epsilon}\,) \nonumber\\
&+&G_4\,i(\vec{p}_p\times\vec{p_K}\cdot\vec{\epsilon}\,) +
\ i(\vec{p}_p\times\vec{p}_K\cdot\vec{q}\,)\;
\left[\,G_5\,(\vec{q}\cdot\vec{\epsilon}\,) \right. \nonumber\\
&+&\left.G_6\,(\vec{p}_p\cdot\vec{\epsilon}\,) +
G_7\,(\vec{p}_K\cdot\vec{\epsilon}\,)\,\right]
+\ G_8\,(\vec{\sigma}\cdot\vec{q}\,) (\vec{q}\cdot\vec{\epsilon}\,) 
 \nonumber\\
&+&G_9\,(\vec{\sigma}\cdot\vec{q}\,)  (\vec{p}_p\cdot\vec{\epsilon}\,) +
G_{10}\,(\vec{\sigma}\cdot\vec{q}\,)  (\vec{p}_K\cdot\vec{\epsilon}\,) 
\nonumber\\
&+&\ G_{11}\,(\vec{\sigma}\cdot\vec{p}_p) (\vec{q}\cdot\vec{\epsilon}\,) +
G_{12}\,(\vec{\sigma}\cdot\vec{p}_p) (\vec{p}_p\cdot\vec{\epsilon}\,)
\nonumber\\
&+&G_{13}\,(\vec{\sigma}\cdot\vec{p}_p) (\vec{p}_K\cdot\vec{\epsilon}\,) + 
\ G_{14}\,(\vec{\sigma}\cdot\vec{p}_K) (\vec{q}\cdot\vec{\epsilon}\,) 
\nonumber\\
&+&G_{15}\,(\vec{\sigma}\cdot\vec{p}_K) (\vec{p}_p\cdot\vec{\epsilon}\,) +
G_{16}\,(\vec{\sigma}\cdot\vec{p}_K) (\vec{p}_K\cdot\vec{\epsilon}\,)\,.
\label{general-amplitude}
\end{eqnarray}
Expressions for the CGLN-like amplitudes $G_j$ in terms of the scalar 
amplitudes $A_j$ and kinematical variables are given in Appendix~\ref{CGNL}. 
In the special case of $\vec{p}_p=0$ one obtains the ordinary amplitude in 
the laboratory frame
\begin{eqnarray}
\vec{J}_{L}\cdot\vec{\epsilon} &=& 
 G_1\,(\vec{\sigma}\cdot\vec{\epsilon}\,) +
 G_3\,i(\vec{p}_K\times\vec{q}\cdot\vec{\epsilon}\,) \\
 &+&G_8\,(\vec{\sigma}\cdot\vec{q}\,)  (\vec{q}\cdot\vec{\epsilon}\,) +
 G_{10}\,(\vec{\sigma}\cdot\vec{q}\,)  (\vec{p}_K\cdot\vec{\epsilon}\,)
 \nonumber\\  
 &+&G_{14}\,(\vec{\sigma}\cdot\vec{p}_K) (\vec{q}\cdot\vec{\epsilon}\,) +
 G_{16}\,(\vec{\sigma}\cdot\vec{p}_K) (\vec{p}_K\cdot\vec{\epsilon}\,)\,,
  \nonumber  
\label{lab-case} 
\end{eqnarray}
which is equivalent to Eq.~($4.3$) of \cite{SF1994} and which was 
used in the DWIA calculations of the hypernuclear cross sections in 
Refs.~\cite{archival,AIP2018}.
The formulas for $G_j$ agree with the corresponding 
formulas for $F_i$ ($i=$1,...6) in Eqs.~(B$\cdot$3) of Ref.~\cite{SF1994}. 
Similarly, we can compare Eq.~(\ref{general-amplitude}) for the center-of-mass 
frame ($\vec{p}_p=-\vec{q}$) with that in Ref.~\cite{SF1994} 
and with corresponding CGLN amplitudes in (B$\cdot$1)~\cite{SF1994}. 

In evaluating the matrix element (\ref{matrix_element}) it is convenient 
to use the spherical form of the amplitude
\begin{eqnarray}
\vec{J}\cdot\vec{\epsilon}&=&
\!-\sqrt{3}\,[\,J^{(1)}\otimes\epsilon^{(1)}\,]^0= 
\sum_{\lambda}(-1)^{-\lambda}\,
J^{(1)} _\lambda\epsilon^{(1)}_{-\lambda}\,,\;\;\;\; 
\label{scalar-product}
\end{eqnarray} 
where the components $J^{(1)}_\lambda$ are defined via 12 spherical 
amplitudes ${\cal F}^S_{\lambda\eta}$ with $S=0, 1$ and $\lambda, \eta = \pm1,0$:  
\begin{equation}
J^{(1)}_\lambda = \sum_{S\eta}\,{\cal F}^S_{\lambda\eta}\,
\sigma^S_\eta\,.
\label{spherical-ampl}
\end{equation}
Here $\sigma^1_\eta$ are the spherical components of Pauli matrices and 
$\sigma^0_0$ is the unit matrix. Inserting (\ref{spherical-ampl}) 
into (\ref{scalar-product}) we get the following explicit form 
\begin{eqnarray} 
\vec{J}\cdot\vec{\epsilon}=&-&\!\!\epsilon^{\;1}_{\;1}\;(\,
{\cal F}^{\;0}_{\!\!-10} +\sigma^1_{\;1}\,{\cal F}^{\;1}_{\!\!-11} +
\sigma^1_0\,{\cal F}^{\;1}_{\!\!-10} 
+\sigma^1_{\!-1}\,{\cal F}^{\;1}_{\!\!-1-1}\,) \nonumber\\
&+&\!\!\epsilon^{\;1}_{\;0}\;(\,{\cal F}^{\,0}_{00}\;\, +
\sigma^1_1\,{\cal F}^{\,1}_{01}\; +\sigma^1_0\,{\cal F}^{\,1}_{00}\; 
+\sigma^1_{\!-1}\,{\cal F}^{\;1}_{0-1}\,) \nonumber\\
&-&\!\!\epsilon^{\;1}_{-1}\,(\,{\cal F}^{\,0}_{10}\; +
\sigma^1_1\,{\cal F}^{\,1}_{11}\; +\sigma^1_0\,{\cal F}^{\,1}_{10}\; 
+\sigma^1_{\!-1}\,{\cal F}^{\;1}_{1-1}\,)\,.\nonumber\\
\label{spherical-form}
\end{eqnarray}
The formulas for ${\cal F}^S_{\lambda\eta}$ written in terms of the 
spherical components of the momenta and the CGLN-like amplitudes $G_j$
are given in Appendix~\ref{spherical}.
%
\subsection{Cross section for hypernuclear production}
The spherical components ($T^{(1)}_\lambda$) of the hypernuclear   
production amplitude (${\cal T}_\mu$) can be decomposed into the 
reduced amplitudes $A^\lambda_{Jm}$ 
\begin{equation}
T^{(1)}_\lambda =\frac{1}{[J_H]}\sum_{Jm}\,{\sf C}^{J_HM_H}_{J_AM_AJm}\,
A^\lambda_{Jm}\,,
\label{amplitude}
\end{equation}
where ${\sf C}^{J_HM_H}_{J_AM_AJm}$ is the Clebsch-Gordan 
coefficient, ($J_A$, $M_A$) and ($J_H$, $M_H$) are the nuclear and hypernuclear  
spin and its projection, respectively, and $[J]=\sqrt{2J+1}$, which is used 
also in the following relations. In the OFA (\ref{matrix-5}) and assuming 
that the proton and $\Lambda$ are in single-particle states $\alpha$ ($=nlj$) 
and $\alpha^\prime$, respectively, we get 
\begin{eqnarray}
A^\lambda_{Jm} =&&\frac{1}{[J]}\,\sum_{S\eta}\,{\cal F}^S_{\lambda\eta}\,
\sum_{LM}\;{\sf C}^{Jm}_{LMS\eta}\sum_{\alpha'\alpha}
{\cal R}^{LM}_{\alpha'\alpha}\;{\cal H}^{LSJ}_{l'j'lj}\nonumber\\
&&\times(\,\Phi_H\,||\,[b_{\alpha'}^+\otimes a_\alpha]^J\,||\;\Phi_A\,)\,,
\label{amplitude-3}
\end{eqnarray}
where ${\cal R}^{LM}_{\alpha'\alpha}$ are the radial integrals, 
${\cal H}^{LSJ}_{l'j'lj}$ includes the Racah algebra (see Eq.~(\ref{HO-matrix-element})), 
and the reduced one-body density matrix elements (OBDME) are non zero 
only for the considered single-particle transitions 
$(\alpha)_p\to(\alpha')_\Lambda$ described by the operator 
$[b_{\alpha'}^+\otimes a_\alpha]^J$ with the proton annihilation 
($a_\alpha$) and $\Lambda$ creation ($b^+_{\alpha'}$) operators.
More details on derivation of Eq.~(\ref{amplitude-3}) is given 
in Appendix~\ref{details}.

%
%
\begin{figure}[htb]
\begin{center}
\epsfig{file=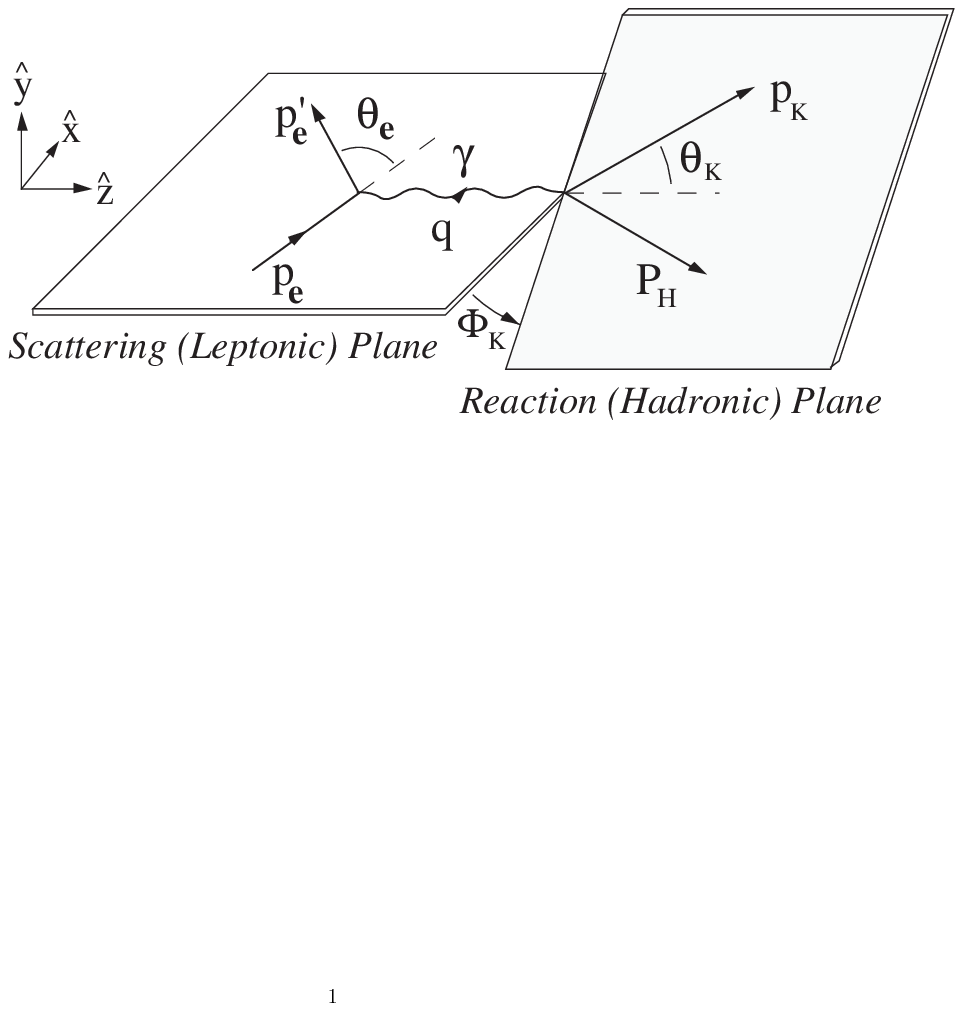,width=7.4cm}\hspace{7mm}
\epsfig{file=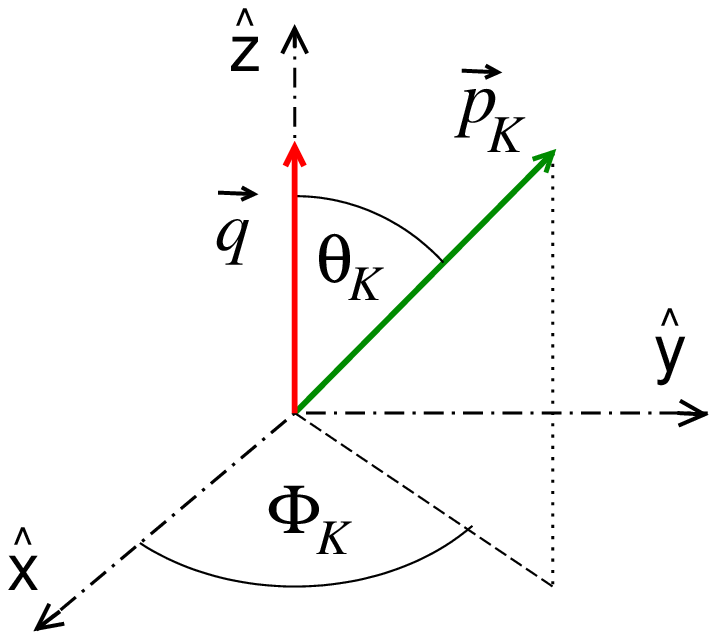,width=4.1cm}
\end{center}
\caption{The laboratory frame for hypernucleus electroproduction 
($\vec{P}_A=0$) with the $z$-axis oriented along the photon momentum $\vec{q}$. 
The $y$-axis of the right-handed system is perpendicular to the leptonic plane, 
see the upper part of the figure. The kaon momentum is defined by its polar 
$\theta_K$ and azimutal $\Phi_K$ angles, see the lower part. Similar notation 
is used for the proton momentum $\vec{p}_p$. The angle $\Phi_K$ also defines 
the angle between the leptonic and hadronic planes. Note our convention that 
for $\Phi_K=0$ the kaon momentum lies between the beam ($\vec{p}_e$) and 
photon momenta. }
\label{frame}
\end{figure}

The cross sections are calculated in the laboratory frame ($\vec{P}_A=0$) 
with electron kinematics and definitions of the angles as given in Fig.~\ref{frame}. 
The virtual photon has the energy $E_\gamma= E_e - E_e^\prime$, a ``mass'' 
$Q^2= -q^2 = \vec{q}^{\;2}-E_\gamma^2$, the transverse polarization 
\begin{equation}
\varepsilon = \left( 1+\frac{2|\vec{q}|^2}{Q^2}\, 
\text{tan}^2\frac{\theta_e}{2} \right)^{-1}\,,
\end{equation}
and the longitudinal polarization given as $\varepsilon_L = \varepsilon\,Q^2/E_\gamma^2$. 
The unpolarized differential cross section for electroproduction is then 
\begin{eqnarray}
\frac{d\sigma}{d\Omega_K} = 
\frac{d\sigma_{\sf T}}{d\Omega_K} +
\varepsilon_L\frac{d\sigma_{\sf L}}{d\Omega_K} +
\varepsilon\frac{d\sigma_{\sf TT}}{d\Omega_K} +
\sqrt{\varepsilon_L(\varepsilon+1)}\frac{d\sigma_{\sf TL}}{d\Omega_K},\nonumber\\
\frac{d^3\sigma}{dE_e^\prime d\Omega_e^\prime d\Omega_K} = 
\Gamma\,\frac{d\sigma}{d\Omega_K}\,,\ \ \ \ \ \ \ \ \ \ \ \ \ \ \ \ \ \ \ \ \  
\label{crs}
\end{eqnarray}  
where the triple-differential cross section is a product of 
$\text{d}\sigma/\text{d}\Omega_K$ and the virtual-photon flux in the 
nucleus-rest frame  
\begin{equation}
\Gamma = \frac{\alpha}{2\pi^2} \frac{|\vec{q}\,|}{Q^2(1-\varepsilon)} 
\frac{E_e^\prime}{E_e}\,,
\label{eq:GammaFactor}
\end{equation}
with the fine-structure constant $\alpha$. The separated cross sections are  
\begin{subequations} 
\begin{align}
\frac{d\sigma_{\sf T}}{d\Omega_K} &= \frac{\beta}{2[J_A]^2}\sum_{Jm}
\frac{1}{[J]^2}\left(\,|A^{+1}_{Jm}|^2 + |A^{-1}_{Jm}|^2\right)\,,
\label{crs1} \\
\frac{d\sigma_{\sf L}}{d\Omega_K} &= \frac{\beta}{[J_A]^2}\sum_{Jm}
\frac{1}{[J]^2}\,|A^0_{Jm}|^2\,,
\label{crs2} \\
\frac{d\sigma_{\sf TT}}{d\Omega_K} &= \frac{\beta}{[J_A]^2}\sum_{Jm}
\frac{1}{[J]^2}\,{\sf Re\,}[A^{+1}_{Jm}A^{-1*}_{Jm}] \,,
\label{crs3} \\
\frac{d\sigma_{\sf TL}}{d\Omega_K} &= \frac{\beta}{[J_A]^2}\sum_{Jm}
\frac{1}{[J]^2}\,{\sf Re\,}[A^{0*}_{Jm}(A^{+1}_{Jm} -A^{-1}_{Jm})] \,,
\label{crs4}
\end{align}
\end{subequations}
where the kinematical factor $\beta$ depends on the kaon momentum~\cite{SF1994}. 
In the kinematics considered here and those of the performed experiments~\cite{archival} 
the transverse part $d\sigma_{\sf T}$ dominates the cross section. The longitudinal 
part $d\sigma_{\sf L}$ gives small contributions but, as we will show, it is sensitive 
to kinematical effects and elementary amplitudes. The TL interference part is 
an important contribution to the full cross section even for quite small photon 
virtualities considered here, $Q^2\approx 0.01$ (GeV/c)$^2$. 
%
\subsection{Optimum on-shell approximation}
\label{optimum}
The elementary electroproduction amplitude is constructed assuming that 
the involved particles are on their mass shell, except the virtual 
photon, and that the energy-momentum is conserved. However, in the impulse 
approximation (Fig.~\ref{impulse-approximation}) the initial proton and 
final $\Lambda$ are not asymptotically free objects, in fact they 
are intermediate particles. In our model calculations we deem that it is 
reasonable to keep the baryons on their mass shell and consider a translational 
invariance of the elementary and overall amplitudes and of the nuclear and 
hypernuclear wave functions, the latter being written as a plane wave for 
the center-of-inertia motion multiplied by the internal part as a function 
of relative (Jacobi) coordinates. We also require energy conservation in 
the elementary vertex 
\begin{equation}
E_\gamma +\sqrt{m_p^2+\vec{p}_p^{\;2}}= 
\sqrt{m_K^2+\vec{p}_K^{\;2}} +\sqrt{m_\Lambda^2+\vec{p}_\Lambda^{\;2}} 
\label{2b-ec}
\end{equation}
to keep the amplitude on the energy shell as well. This requirement together 
with energy conservation in many-body vertexes 
\begin{subequations}
\begin{align}
E_A &= \sqrt{M_c^2+(\vec{P}_A-\vec{p}_p)^2} 
+\sqrt{m_p^2+\vec{p}_p^{\;2}}+\epsilon_p\,,
\label{2b-vertex} \\
E_H &= \sqrt{M_c^2+(\vec{P}_A-\vec{p}_p)^2} 
+\sqrt{m_\Lambda^2+\vec{p}_\Lambda^{\,2}}+\epsilon_\Lambda\,,
\label{mb-vertex}
\end{align}
\end{subequations}
where $\epsilon_p$ and $\epsilon_\Lambda$ are single-particle binding 
energies in a general reference frame, leads to a violation of the 
overall energy conservation by a factor $\epsilon_p-\epsilon_\Lambda$. 
However, as this difference is relatively small, 
$\epsilon_p-\epsilon_\Lambda \approx 10$ MeV, in comparison with the total 
energy, $E_\gamma +E_A \approx 10$ GeV, we neglect it assuming the overall 
energy conservation  
\begin{equation}
E_\gamma +E_A = E_K +E_H\,.
\label{mb-ec}
\end{equation}

The kinematics of electroproduction is determined by the virtual-photon 
momentum $q$ and polarization $\varepsilon$ and by the kaon polar $\theta_K$ 
and azimuthal $\Phi_K$ angles. The magnitude of the kaon three-momentum 
has to be calculated from energy conservation Eqs.~(\ref{2b-ec}) or 
(\ref{mb-ec}), which we denote by $|\vec{p}_K(2b)|$ and $|\vec{p}_K(mb)|$, 
respectively. Since the solution $|\vec{p}_K(2b)|$ for the 2-body reaction 
with an arbitrary proton momentum in general differs from the solution 
$|\vec{p}_K(mb)|$ for the many-body process, we have to choose which value 
will be used in computing the elementary amplitude (ea) and the radial 
integral (ri) in Eq.~(\ref{amplitude-3}) and for the kinematic factor 
$\beta$ in Eqs.~(\ref{crs1})--(\ref{crs4}). 
This provides us with various schemes of calculation:\\
(a) {\bf 2-body}: only one value $|\vec{p}_K(2b)|$ is used for ea, ri, 
     and $\beta$. The elementary amplitude is then on-energy-shell 
     but the many body energy conservation (\ref{mb-ec}) is violated.\\
(b) {\bf 2-body\_ea}: a hybrid scheme with {\it two different values} 
    $|\vec{p}_K(2b)|$ for ea and $|\vec{p}_K(mb)|$ for ri and $\beta$.
    The elementary amplitude is on-energy-shell and both (\ref{2b-ec}) 
    and (\ref{mb-ec}) are fulfilled. Note that this scheme was used 
    in our previous calculations~\cite{archival,AIP2018}.\\
(c) {\bf many-body}: only one value $|\vec{p}_K(mb)|$ is used for ea, ri, 
     and $\beta$. However, in this case the elementary amplitude is 
     off-energy-shell 
     which causes additional uncertainty of the results.\\
In the next section we will show differences of the cross sections 
calculated in these schemes.  
     
One can take advantage of the possibility to choose the effective value of 
$\vec{p}_p$ in Eq.~(\ref{2b-ec}) and find a value that gives the same 
solution as that of Eq.~(\ref{mb-ec}), $|\vec{p}_K(2b)|=|\vec{p}_K(mb)|$. 
This value exists and we denote it as an ``optimum'' proton 
momentum $\vec{p}_\text{opt}$. In fact, we have one equation for the magnitude 
$|\vec{p}_\text{opt}|$ and an angle with respect to $\vec{\Delta}$ which 
can be chosen. In our calculations we have chosen this angle to be 
180$^\circ$, {\it i.e.}, the proton is moving opposite to the momentum 
transfer minimizing the momentum of $\Lambda$.    
Note that in the OFA, Eq.~(\ref{matrix-5}), this optimum momentum equals 
$\vec{p}_\text{eff}$ and makes the three schemes given above equivalent 
allowing for the on-energy-shell elementary amplitude. Therefore we denote 
this as the optimum on-shell approximation. 
Note also that we do not perform an ``optimal Fermi averaging'' 
as in Ref.~\cite{Harada} for the ($\pi^+$, K$^+$) production because we  
use the OFA. 
%
\subsection{Mean proton momentum in target nucleus}
\label{mean}
The effective value of the proton momentum in the OFA (\ref{matrix-5}) 
can be also chosen as a mean momentum of the proton determined from 
its mean kinetic energy inside a nucleus, 
$|\vec{p}_\text{eff}| = \sqrt{2 \mu\,\langle E_{kin} \rangle}$,  
where $\mu$ is the reduced mass of the proton-core system. 
In the analysis presented here we calculate the mean kinetic energy 
of a proton inside $^{12}$C using the single-particle wave function 
of the proton used also for computing the radial integrals. 
The interaction between the proton and the core $^{11}$B is described 
by the Woods-Saxon and Coulomb potential which was also used in 
Ref.~\cite{archival}. For the protons bound in the $0p_{1/2}$ 
and $0p_{3/2}$ states with the binding energies $-10.37$ and 
$-15.96$ MeV we have obtained the mean kinetic energies 18.15~MeV 
and 18.76~MeV, respectively. These very near values, 
even if the binding energies are quite different, can be attributed 
to a relatively strong spin-orbital part of the potential with 
a depth of $V_{LS}=-19$ MeV. The mean momenta are then  
176.7 and 179.6 MeV/c. In our comparison presented in the next 
section we will consider that $|\vec{p}_\text{eff}|= 179$ MeV/c 
with two values of the angle with respect to the photon, 
$\theta_\text{eff}= 0^\circ$ and 180$^\circ$.      
%
%
%
\section{Results}
First, we give more details about the model calculations presented here 
mentioning also upgrades of the model with respect to the previous version 
and then we show kinematic and Fermi motion effects on the angular 
($\theta_{Ke}$) and energy ($E_\gamma$) dependence of the cross sections 
in electroproduction of $^{12}_{~\Lambda}$B. We prefer to use the kaon 
angle with respect to the beam, $\theta_{Ke}$, instead of with respect to 
the photon, $\theta_K$, see Fig.~\ref{frame}. The results are presented 
only for selected hypernuclear states ($E_x$[MeV], ${\sf J^P}$): (0.0, 1$^-$) 
and (0.116, 2$^-$) with $\Lambda$ in $s$ orbit and (10.525, 1$^+$), 
(11.059, 2$^+$), and (11.132, 3$^+$) with $\Lambda$ in $p$ orbit. 
It is interesting to mention a selection rule according to which 
contributions from the spin non-flip part of the elementary amplitude 
${\cal F}^0_{\lambda 0}$ in Eq.~(\ref{amplitude-3}) are only possible for 
the states $1^-$ and $2^+$ but they are missing in  $2^-$, $1^+$, and $3^+$. 
However, in the kinematic region considered here, {\it i.e.} small 
kaon-photon angles $\theta_K$, the strength of the spin non-flip 
spherical amplitudes ${\cal F}^0_{\lambda 0}$ is very small and 
therefore one cannot expect big differences due to this selection rule. 
There is, however, another ``dynamical'' selection rule which will 
be mentioned below and which generates differences between the results 
for these two groups of states.
%
\subsection{More details of the calculation}
\label{calculations}
The elementary amplitude ${\cal F}^S_{\lambda\eta}$ used in eq.~(\ref{amplitude-3}) 
is described by our recent BS3 isobar model~\cite{SB18} and by the older 
SLA Saclay-Lyon model~\cite{SLA}. Note that SLA does not describe the new 
electroproduction data as well as BS3~\cite{SB18} mainly because these data 
were not available during its construction. 
The OBDME are taken from shell-model structure calculations by John Millener 
with a $\Lambda$N effective interaction~\cite{JohnM} similarly to 
our previous calculations~\cite{archival, AIP2018}. The kaon distortion 
included in the radial integrals is described in the eikonal approximation 
with the first-order optical potential constructed from the separable KN 
amplitude as in Ref.~\cite{archival}.

A new feature of the present calculations is that the radial integrals 
${\cal R}^{LM}_{\alpha^\prime\alpha}$ are calculated with proper 
single-particle wave functions according to the quantum numbers 
$\alpha$ and $\alpha^\prime$ of assumed transitions given by OBDME. 
In the previous calculations we used only the radial integrals 
with the quantum numbers of dominant transitions. 
Another improvement is using the right relative particle-core coordinate 
described by the Jacobi coordinate $|\vec{\xi}|$ in the radial integrals. 
This allows one to calculate the proton and $\Lambda$ wave functions 
from the Schr\"odinger equation with a particle-core potential. 
Here we use the Woods-Saxon and Coulomb potentials as in~\cite{archival}.
Note also that when checking the new computer code we found that in 
our old code there was a wrong sign at the Clebsch-Gordan coefficient 
with $J=3$ which significantly changed results for the corresponding 
states. We have also corrected for a tiny flaw in calculating the 
virtual-photon flux for hypernucleus electroproduction. 
These upgrades of the calculations make some differences which 
will be shown and discussed in Fig.~\ref{scheemes} and Table~\ref{TabBS3}.   

The calculations are performed in the coplanar kinematics ($\Phi_K$ = 180$^\circ$), 
see Fig.~\ref{frame}, 
close to that of the Hall A experiment~\cite{archival} with 
$Q^2$ = 0.06 (GeV/c)$^2$, $\varepsilon$ = 0.7, $E_\gamma\in(1.5, 2.5)$ GeV, 
and $\theta_{Ke}\in(5^\circ, 14^\circ)$.
    
\subsection{Kinematic and Fermi motion effects in the cross sections}
%
%
%
\begin{figure*}[ht!]
\begin{center}
\includegraphics[width=0.43\textwidth,angle=270]{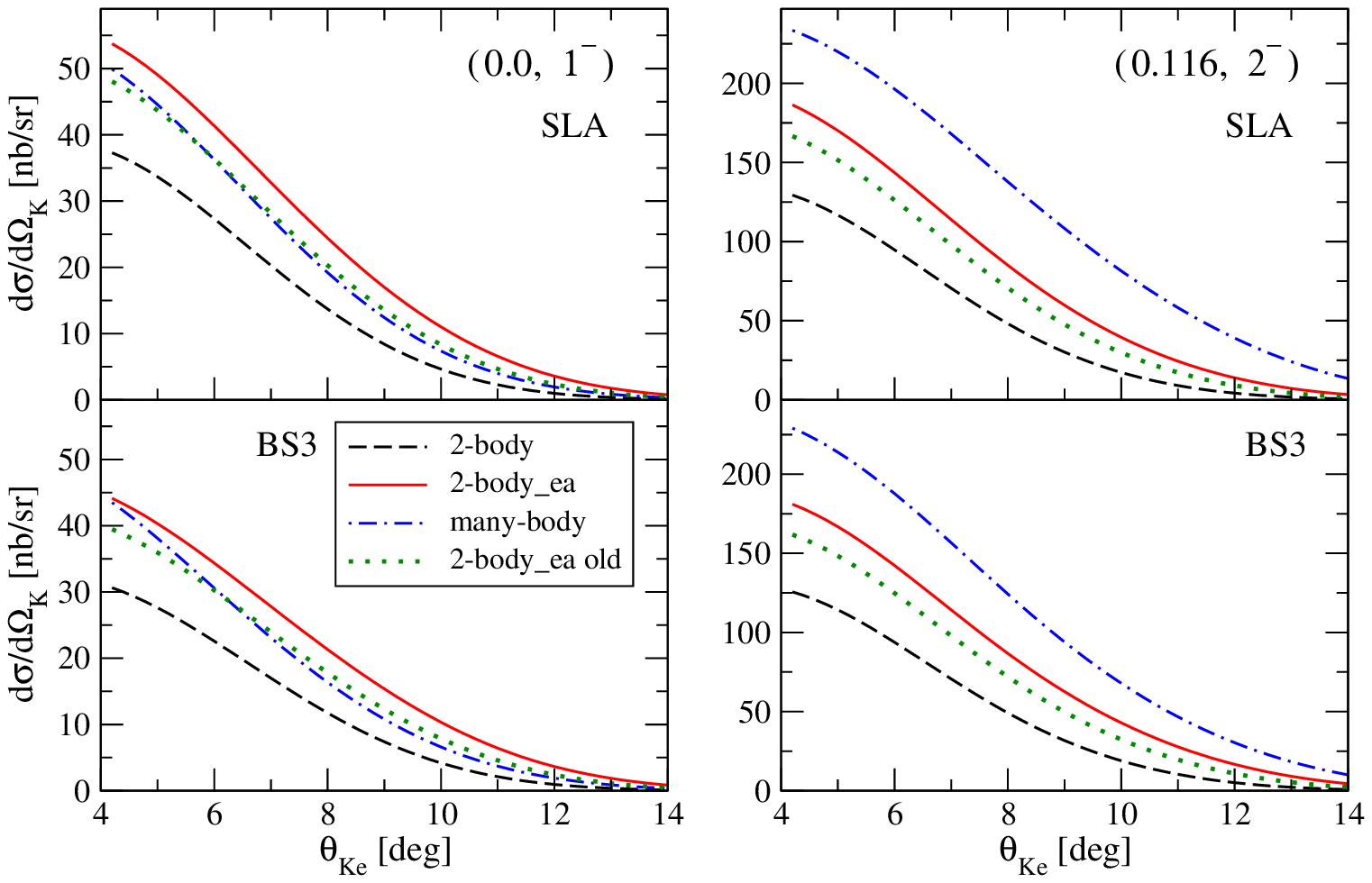}
\includegraphics[width=0.52\textwidth,angle=270]{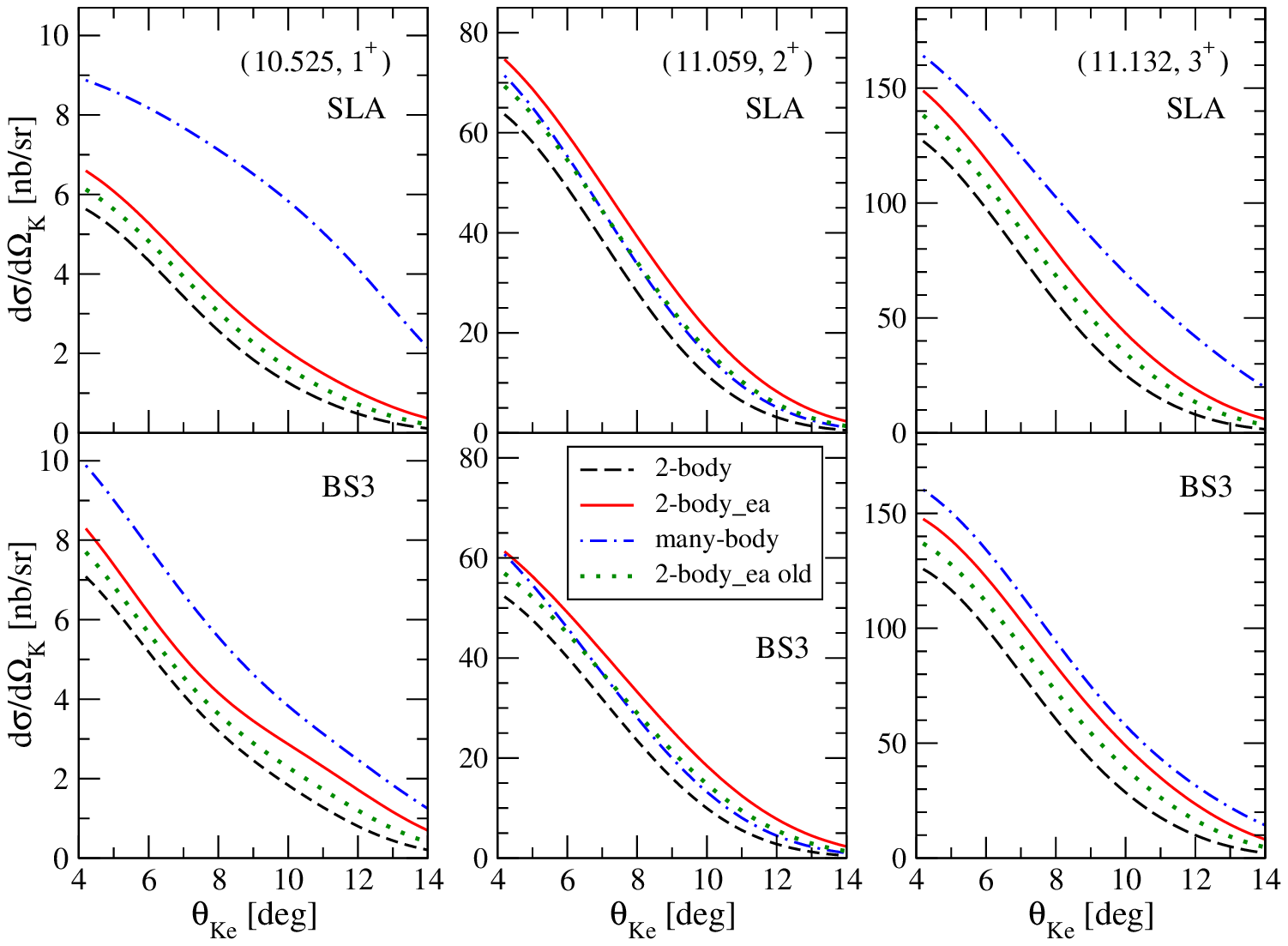}
\end{center}
\caption{The full electroproduction cross section for the selected 
states of $^{12}_{~\Lambda}$B calculated in the frosen-proton 
approximation ($\vec{p}_{\sf eff}$ = 0) with the SLA and BS3 amplitudes and 
using various kaon momenta (computational schems).   
The results denoted as ``2-body\_ea old'' are calculated 
without the upgrades in the radial integrals (see Sec.~\ref{calculations}) 
and therefore can be compared with our previous calculations in~\cite{archival}.}
\label{scheemes}
\end{figure*} 
Kinematic effects caused by using different kaon momenta in schemes 
2-body, 2-body\_ea, and many-body are shown in Fig.~\ref{scheemes} on the 
angular dependence of the full cross section $d\sigma / d\Omega_K$. 
The calculations are in the frozen-proton approximation ($\vec{p}_{\sf eff}=0$) 
with the photon energy $E_\gamma= 2.2$ GeV. The upper and lower parts of the panels 
show results with the SLA and BS3 amplitudes, respectively, and each panel is for 
a given hypernuclear state. 
The result ``2-body\_ea old'' was calculated with ``absolute'' coordinates in 
the radial integrals and with the radial integrals for the dominant transition 
($\alpha\rightarrow\alpha^\prime$) as in our previous calculations~\cite{archival}. 
The absolute coordinates are suitable when we use the harmonic-oscillator 
single-particle wave functions but  
in calculations with the Woods-Saxon wave functions the relative particle-core 
(Jacobi) coordinates are more suitable. 
Comparison of the results ``2-body\_ea'' and ``2-body\_ea old'' therefore shows 
changes due to improvements in our model calculations. The new results are larger by 
a few per-cent. 

The effects, given by different values of the kaon momentum $|\vec{p}_K|$ 
used in various components of the cross section, are quite large and do not 
depend much on the elementary amplitude. They are even larger in some 
cases for SLA than for BS3, see {\it e.g.} the results for the $1^+$ and 
$3^+$ states. Even if the difference between $|\vec{p}_K(2b)|$ and  
$|\vec{p}_K(mb)|$ in the considered kinematics and 
$\theta_{Ke}$ = 6$^\circ$ is only smaller than 2\%, the difference 
of corresponding momentum transfer $|\vec{\Delta}|$ is about 10\%, which raises 
values of the radial integrals. 
Comparison of the results in the 2-body and 2-body\_ea schemes show effects 
from the radial integrals and also from the normalization of the cross sections 
by the parameter $\beta$. As the angular dependence of $\beta$ is weak and 
$\beta[\vec{p}_K(mb)] > \beta[\vec{p}_K(2b)]$ the effect from 
$\beta$ is mainly the angle-independent rescaling of the cross section,  
which is observed for both elementary amplitudes and all states in 
Fig.~\ref{scheemes}. One can therefore conclude that even a small violation 
of the many-body energy conservation in the 2-body scheme makes the cross 
sections smaller by 10--20\% for all states.  

The effect of using the off-energy-shell elementary amplitude is seen 
from the comparison of the 2-body\_ea (on-shell) and many-body (off-shell) 
schemes. In some cases, see {\it e.g.} the states $2^-$ and $1^+$ 
in Fig.~\ref{scheemes}, the effect is quite large, amounting to about 50\% 
for $2^-$ at $\theta_{Ke}=6^\circ$, and, of-course, it depends on 
the amplitude. In general, bigger differences between the results are 
observed for the $2^-$, $1^+$, and $3^+$ states whereas the effects 
for the other group of states, $1^-$ and $2^+$ are smaller. 
Interestingly, the curves for the two groups of states, 
see \textit{e.g.} states $1^-$ and $2^-$, are ordered in a different way.  
This feature can be understood from a numerical analysis of contributions 
to the reduced amplitude $A^\lambda_{Jm}$ in Eq.~(\ref{amplitude-3}). 
Indeed, from the analysis one can conclude that the radial 
integrals with $M=0$ acquire the largest values, particularly 
their imaginary parts, and are rising functions of $|\vec{p}_K|$. 
Note also that $|\vec{p}_K(mb)| > |\vec{p}_K(2b)|$.  
Another observation is that the elementary amplitude ${\cal F}^1_{00}$  
dominates in both 2-body\_ea and many-body schemes.
This dominance of contributions with $M=0$ and $\eta=0$ provides 
a selection rule given by values of the Clebsch-Gordan 
coefficient ${\sf C}^{J0}_{L010}$ in Eq.~(\ref{amplitude-3}). 
Recall also that $J=J_H$ as $J_A=0$ for the ground state of $^{12}$C. 
This dynamical selection rule induces a dominance of the longitudinal 
amplitude $A^0_{20}$ for the $2^-$ state and therefore also 
a significant enhancement of the longitudinal (\ref{crs2}) 
and interference (\ref{crs4}) cross section.  
The large values of $d\sigma_{\sf L}$ and $d\sigma_{\sf TL}$ add up 
with $d\sigma_{\sf T}$ giving enhancement of the full cross section 
in the many-body scheme for the $2^-$ state for both SLA and BS3. 
On the other hand, in the $1^-$ state the amplitude $A^0_{10}=0$ 
due to the selection rule ${\sf C}^{10}_{1010}=0$ and the many-body 
cross section is even smaller than that in the 2-body\_ea scheme. 
In this case the cross section 
is only given by the transverse part of the amplitude $A^{\pm 1}_{1m}$.

%
%
\begin{table*}[hbt!]
\begin{center}
\caption{Numerical results calculated with $Q^2=0.06$ (GeV/c)$^2$, $E_\gamma = 2.2$ 
GeV, $\epsilon = 0.7$, $\theta_{Ke}= 6^\circ$, and $\Phi_K= 180^\circ$ are shown for 
three computational schemes 2-body (2b), 2-body\_ea (2b\_ea), and many-body (mb). 
The full cross section (d$\sigma$) and contributions from the transverse (T), 
longitudinal (L), and interference (TL) parts are shown together with the reduced 
amplitudes with $\lambda = 0$ and dominant components of the elementary amplitude BS3. 
The cross sections are in nb/sr. The bold values indicate the longitudinal 
contributions that constitute the selection rule.}
\begin{tabular}{cccccccccc}
\hline\\ 
Scheme &  d$\sigma$ & T & L & TL & $|A^0_{J\pm 1}|$ & $|A^0_{J0}|$ 
&${\cal F}^1_{00}$ &${\cal F}^1_{01}=-{\cal F}^1_{0-1}$\\
\hline\vspace{-3mm}\\
 & & &\multicolumn{4}{c}{State $E = 0.0$ MeV, $J^P = 1^-$} & & & \\
2b     &  22.40 & 25.88 & 0.55 & -4.21 & 0.165 & 0.0 & (0.911, 2.529) &   
(-0.535, 0.203)  \\
2b\_ea &  34.12 & 39.18 & 0.93 & -6.24 & 0.213 & 0.0 & (0.911, 2.529) &   
(-0.535, 0.203)  \\
mb     &  30.34 & 38.75 & 1.48 &-10.10 & 0.270 & 0.0 & (-3.607, 3.114) &   
 (0.108, 0.262) \\
\hline\vspace{-3mm}\\
 & & &\multicolumn{4}{c}{State $E = 0.116$ MeV, $J^P = 2^-$} & & & \\
2b     &  93.80 &  89.74 & {\bf 17.18} & -13.71 & 0.364 & {\bf 1.610} &
 (0.911, 2.529) &  (-0.535, 0.203)  \\
2b\_ea & 142.32 & 135.87 & {\bf 25.82} & -20.26 & 0.461 & {\bf 1.947} &
 (0.911, 2.529) &   (-0.535, 0.203)  \\
mb     & 187.77 & 134.46 & {\bf 77.59} & -25.04 & 0.571 & {\bf 3.474} &
 (-3.607, 3.114) &  (0.108, 0.262) \\
\hline
\end{tabular}
\label{2b-mb}
\end{center}
\end{table*}
In Table~\ref{2b-mb} we show numerical results for the 
cross sections, reduced amplitudes, and elementary amplitudes calculated 
with $\theta_{Ke}=6^\circ$ for the states $1^-$ and $2^-$. 
One can see a significant enhancement of $d\sigma_{\sf L}$ and 
$d\sigma_{\sf TL}$ for $2^-$ made by contributions from $|A^0_{20}|$. 
Note also a big value of the off-energy-shell elementary amplitude 
${\cal F}^1_{00}$ in the many-body scheme. 
Similarly, the longitudinal cross section is enhanced 
in the states $1^+$ and $3^+$.  
From the dominance of the off-energy-shell elementary amplitude ${\cal F}^1_{00}$ 
one can also conclude, that the largest off-shell effects will be observed 
in $d\sigma_\text{L}$ and $d\sigma_\text{TL}$ as seen in Table~\ref{2b-mb}. 
As the off-energy-shell extension of the elementary amplitude is barely under 
control we prefer the schemes 2-body and 2-body\_ea with 
the on-energy-shell amplitude. Recall that if we opt for the optimum 
proton momentum, all schemes are equivalent and the elementary 
amplitude is on-energy-shell. 

In the following we will study effects of the proton motion in the target 
nucleus, which we denote as Fermi motion effects. Particularly, we will 
demonstrate these effects on the angle and energy dependent cross sections 
calculated in the OFA with various effective proton momenta. 
We consider five cases:\vspace{1mm}\\ 
\begin{tabular}{rll}
i)  & {\bf frozen p}        & $\vec{p}_{\sf eff}= 0$ 
      $\Rightarrow \vec{p}_\Lambda = \vec{\Delta}$\\
ii) & {\bf frozen $\Lambda$}& $\vec{p}_{\sf eff}= -\vec{\Delta}$ 
      $\Rightarrow \vec{p}_\Lambda = 0$\\
iii)& {\bf optimum}         & $\vec{p}_{\sf eff}= \vec{p}_{\sf opt}$\\  
iv) & {\bf mean 0}          & $|\vec{p}_{\sf eff}|= 179$ MeV/c 
      and $\theta_{p} = 0^\circ$\\
v)  & {\bf mean 180}        & $|\vec{p}_{\sf eff}|= 179$ MeV/c 
      and $\theta_{p} = 180^\circ$\vspace{1mm}\\
\end{tabular}\\
Note that calculations in the frozen $\Lambda$ approximation were 
also performed in Ref.~\cite{HLee} for the 
$^{12}$C($\gamma$,K$^+$)$^{12}_{~\Lambda}$B reaction.  
The optimum and mean values are calculated as described 
in Secs.~\ref{optimum} and \ref{mean}, respectively.
We present the full cross section $d\sigma$ in Eq.~(\ref{crs})  
which corresponds to hypernuclear photoproduction induced by virtual photons. 
In the case of angular dependence the effects observed in $d\sigma$ 
qualitatively coincide with those in the triple-differential cross section 
$d^3\sigma$ because the virtual-photon flux factor $\Gamma$, 
Eq.~(\ref{eq:GammaFactor}), is kaon-angle independent and acts only 
as a scaling factor. 
In the case of the energy dependence the flux factor can modify the shape 
of the curves but as it depends only on the electron kinematics it does 
not qualitatively change effects due to different 
proton and kaon momenta.    

%
%
\begin{figure*}[ht]
\begin{center}
\includegraphics[width=0.44\textwidth,angle=270]{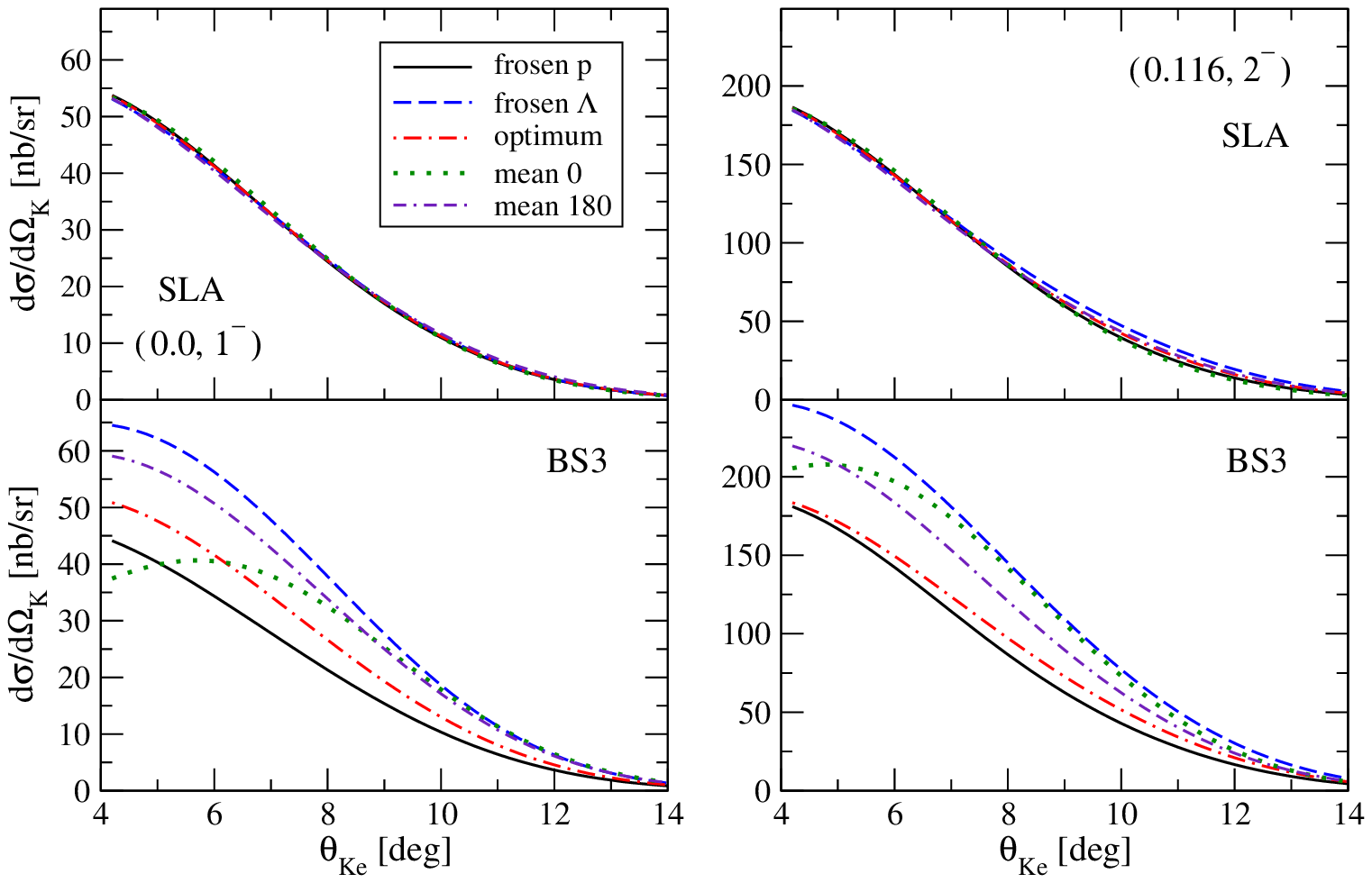}
\includegraphics[width=0.53\textwidth,angle=270]{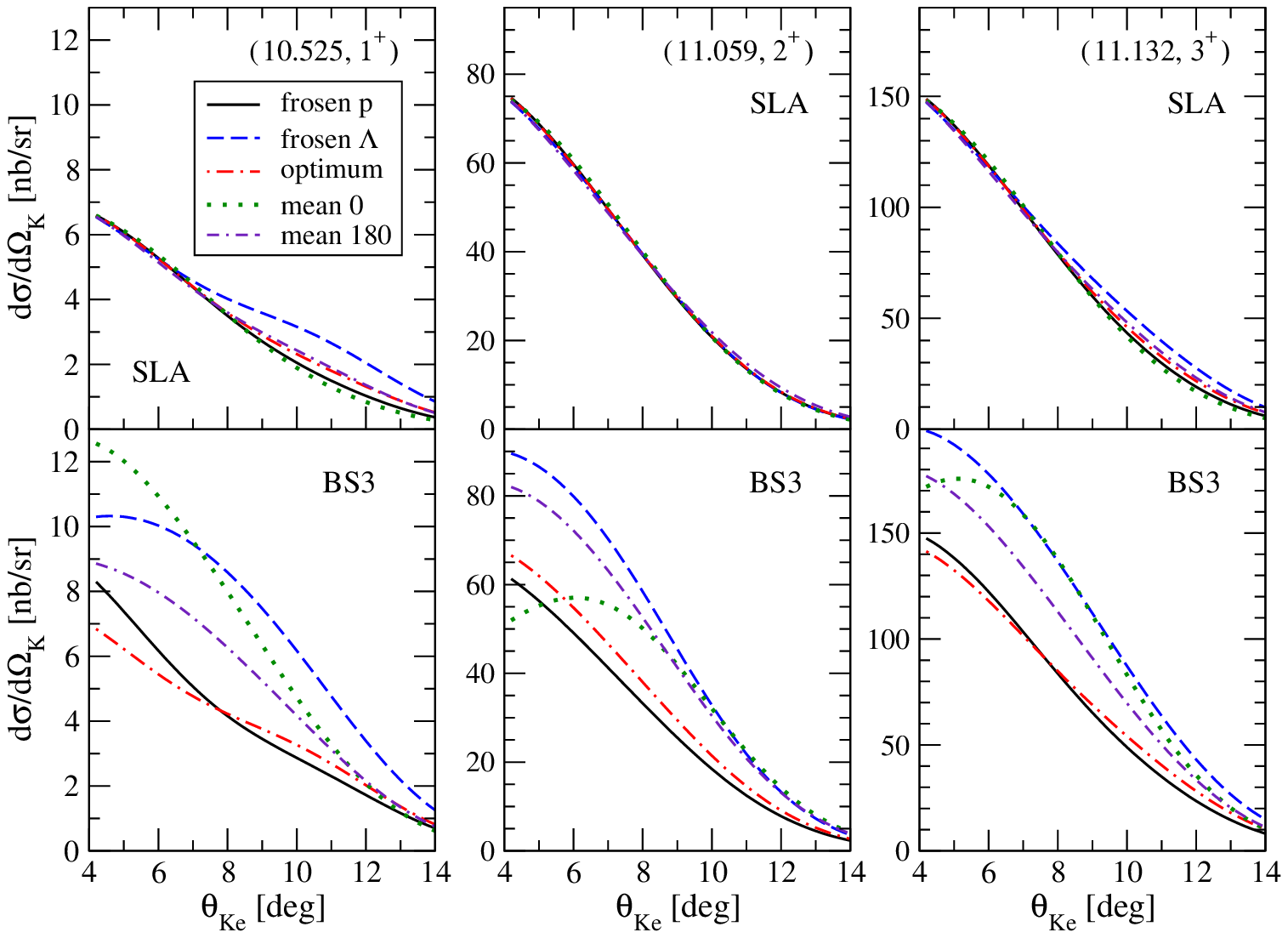}
\end{center}
\caption{The full electroproduction cross section for the selected 
states of $^{12}_{~\Lambda}$B calculated in the scheme 2-body\_ea 
with the SLA nad BS3 elementary amplitudes and with various values 
of the proton momentum $\vec{p}_{\sf eff}$. The photon has energy 
$E_\gamma$= 2.2 GeV, $Q^2$= 0.06 (GeV/c)$^2$, and polarization 
$\varepsilon$ = 0.7.} 
\label{adep-full}
\end{figure*}
The results calculated in the 2-body\_ea scheme with $E_\gamma= 2.2$ GeV 
and with various proton momenta are shown in Fig.~\ref{adep-full}.
The effects from the Fermi motion are very different, both in the 
magnitude and the shape, for the elementary amplitudes SLA and BS3. 
Whereas there are very small effects for the SLA (at given energy), 
the results with BS3 reveal quite a strong dependence on the proton 
momentum. The largest cross sections for a given state are obtained  
with the largest value of the momentum 
$|\vec{p}_{\sf eff}|= |\vec{\Delta}|\approx 300$ MeV/c in the case 
of frozen $\Lambda$, and the smallest values are with frozen proton, 
$|\vec{p}_{\sf eff}|= 0$. These differences at small kaon angles, 
$\theta_K\approx 0$, make about 30\% for BS3. 
Note that the BS3 results with the mean momentum strongly 
depend on the direction with respect to the photon momentum 
($0^\circ$ or $180^\circ$) and that they reveal effects due to 
the dynamical selection rule. This angle dependence differs 
for the SLA nad BS3 amplitudes and it is more pronounced in 
the longitudinal components of the cross section as we will show below. 
The dependence of the effects on the elementary amplitude is also apparent  
from the result with the mean momentum and $\theta_{p}= 0^\circ$ 
(the green dotted line) which lies mostly above the result with frozen 
proton (the black solid line) for the BS3 amplitude 
whereas it is slightly below the black line for the SLA amplitude. 
These effects survive for smaller energies ($E_\gamma= 1.5$ GeV) where 
the Fermi motion effects with SLA are bigger.  
These observations suggest that the effects due to the proton motion 
depend on both the elementary amplitude and kinematics of the process. 

%
%
\begin{figure*}[ht!]
\begin{center}
\includegraphics[width=0.43\textwidth,angle=270]{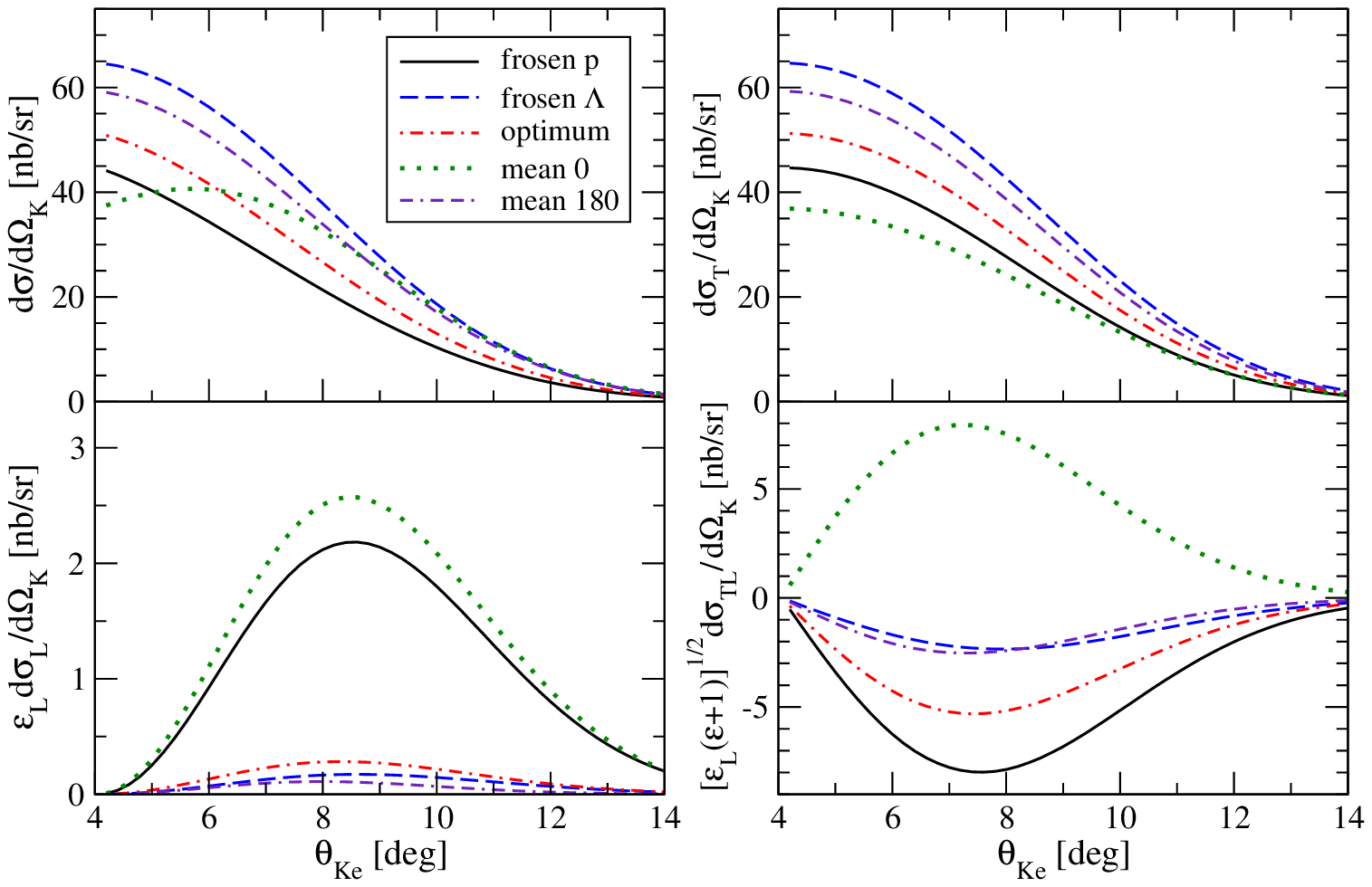}
\includegraphics[width=0.43\textwidth,angle=270]{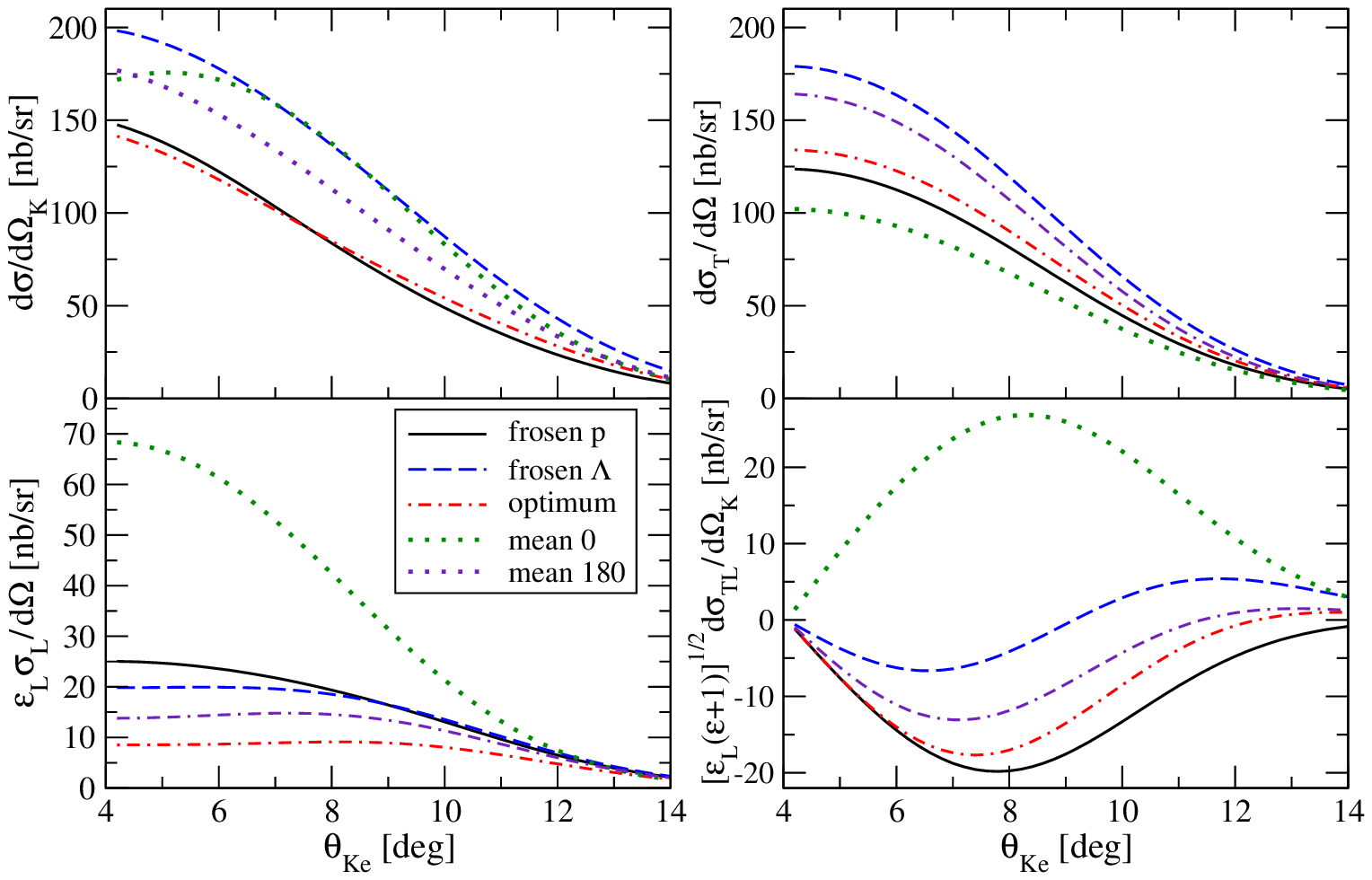}
\end{center}
\caption{Contributions from the transverse (T), longitudinal (L), and 
transverse-longitudinal interference (TL) parts to the full cross section  
in Eq.~(\ref{crs}) are shown for the ground state (upper panels) and 
the $3^+$ excited state (bottom panels) of $^{12}_{~\Lambda}$B. 
The results are calculated in the scheme 2-body\_ea with 
the BS3 amplitude for the five values of the proton momentum 
and kinematics as in Fig.~\ref{adep-full}.}
\label{adep-sep}
\end{figure*}
More pronounced effects of Fermi motion are observed in separated 
contributions from the transverse (T), longitudinal (L), and 
interference (TL) parts of the full cross section as it is 
shown in Fig.~\ref{adep-sep} for the BS3 amplitude and for 
the $1^-$ and $3^+$ states. 
Recall that the results for $1^-$ reveal similar features as 
those for $2^+$ and the results for $3^+$ are similar to those 
for $2^-$ and $1^+$. 
Figure \ref{adep-sep} also shows a difference between the electro- 
and photoproduction calculations represented by the full $d\sigma$ 
for the former and by the transverse $d\sigma_T$ for the latter.  
The transverse part dominates the full cross section in the considered 
kinematic region (small $\theta_{\sf K}$ and $Q^2$), 
but the contributions from $d\sigma_L$ and $d\sigma_{TL}$ are also important 
corrections here, amounting to about 10\% at 6--10 degrees. 
The particular contribution depends on the elementary amplitude. 
Note that the BS3 amplitude includes additional longitudinal couplings 
of the nucleon resonances to the virtual photon~\cite{SB18} which are missing 
in the SLA amplitude. However, the results for the L and TL parts of 
the cross section calculated with SLA also reveal considerable effects 
due to the Fermi motion but these contributions largely cancel each other 
giving only tiny effects in the full electroproduction cross section. 

Whereas the character of the results in Fig.~\ref{adep-sep} for 
the T contribution, the shape and ordering of the lines, is similar 
for both states, the magnitude and behaviour of the L contribution 
are strongly determined by the selection rule and nuclear structure.
The interference part TL being also an important component of the full cross 
section reveals some differences for the presented states as well.
Therefore, it is evident that the different character of the effects in 
the full cross sections for the $1^-$ and $3^+$ states is driven by 
the longitudinal component of the virtual photon. Recall that the 
longitudinal components of $d\sigma$ depend on the reduced amplitude 
$A^0_{Jm}$ with the largest value for $m=0$. Because also the longitudinal 
spherical amplitude ${\cal F}^1_{00}$ dominates (in this kinematic region), 
the behavior of $d\sigma_L$ is strongly affected by the selection rule 
for ${\sf C}^{J_H0}_{L010}$, {\it e.g.} ${\sf C}^{10}_{1010}=0\ $ and 
${\sf C}^{30}_{2010}=\sqrt{3/5}\ $ for the states $1^-$ and $3^+$, respectively. 

\begin{figure*}[htb!]
\begin{center}
\includegraphics[width=0.45\textwidth,angle=270]{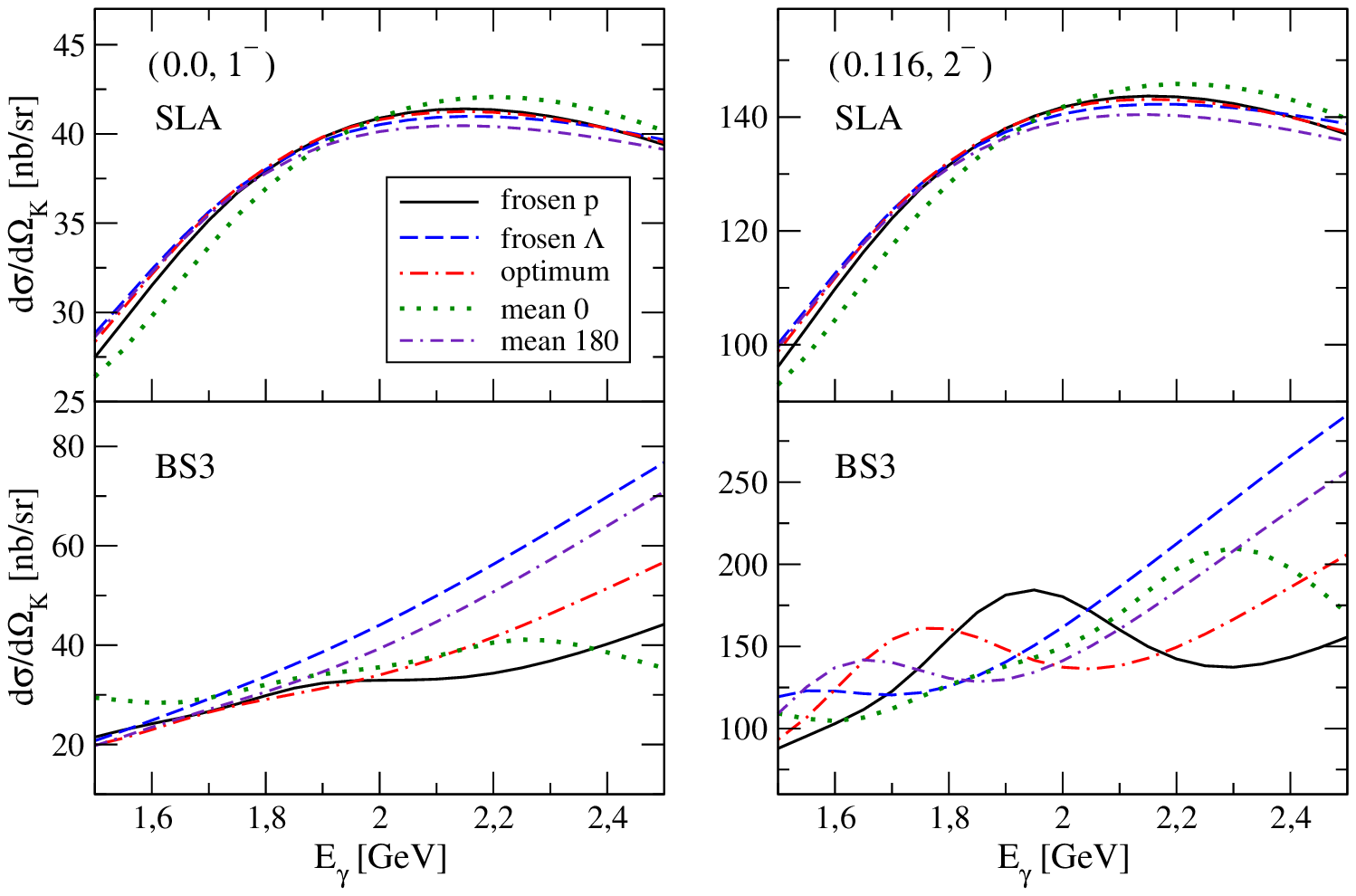}\vspace{1mm}
\includegraphics[width=0.54\textwidth,angle=270]{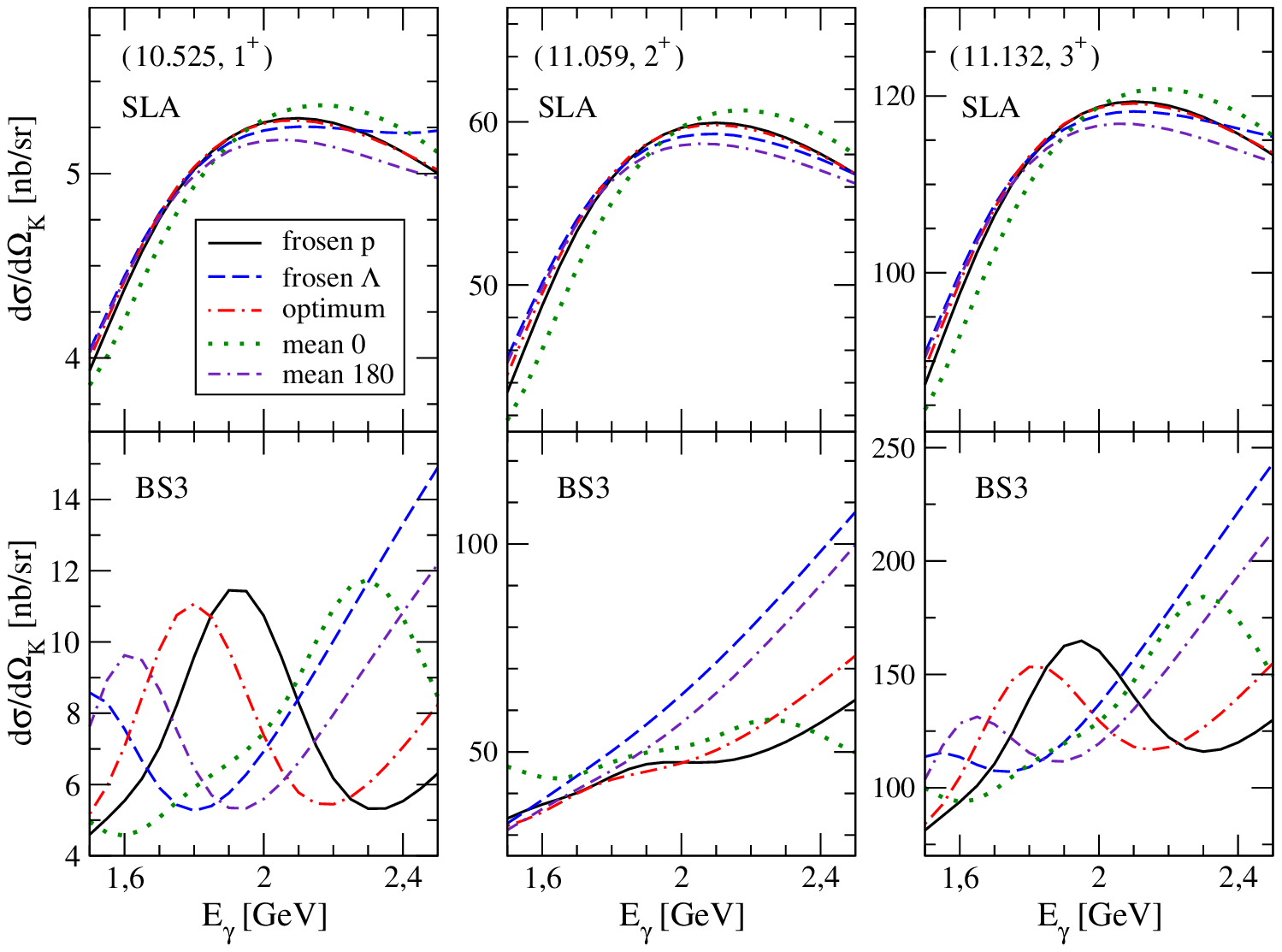}
\end{center}
\caption{The same as in Fig.~\ref{adep-full} but for energy dependence 
at kaon angle $\theta_{Ke}$ = 6$^\circ$.}
\label{edep-full}
\end{figure*}

\begin{figure*}[htb!]
\begin{center}
\includegraphics[width=0.44\textwidth,angle=270]{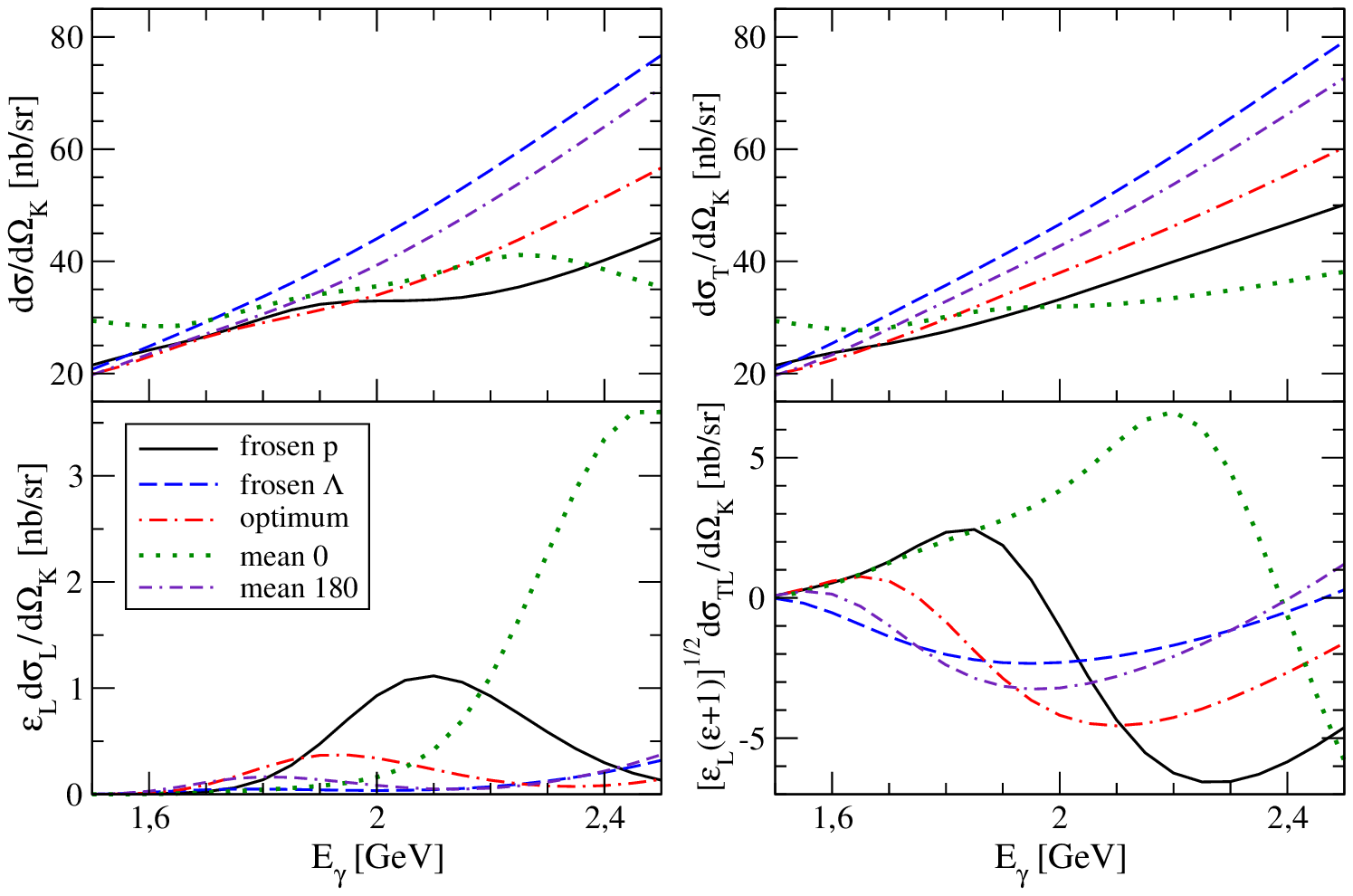}\vspace{1mm}
\includegraphics[width=0.44\textwidth,angle=270]{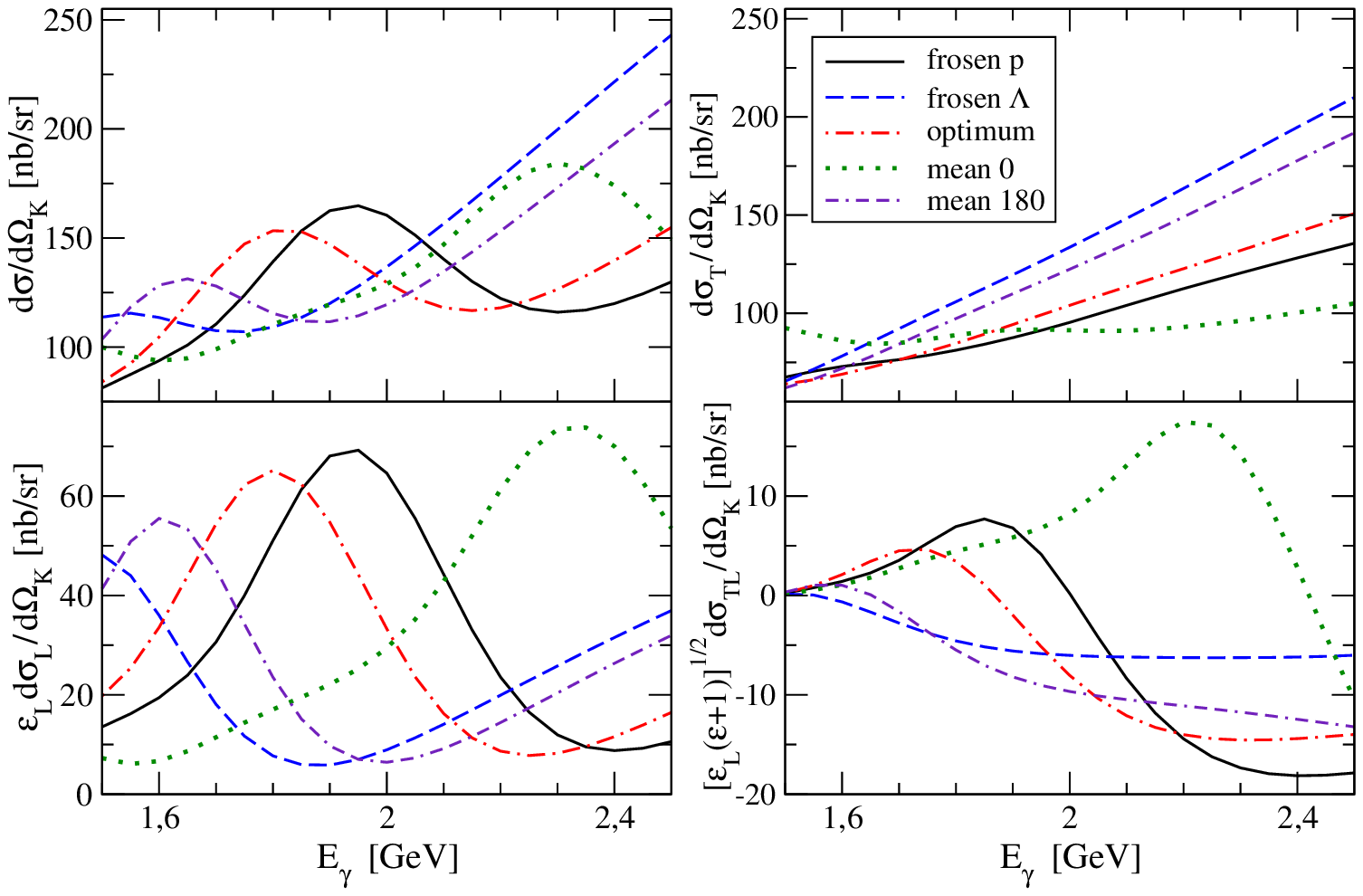}
\end{center}
\caption{The same as in Fig.~\ref{adep-sep} but for energy dependence 
at kaon angle $\theta_{Ke}$ = 6$^\circ$.}
\label{edep-sep}
\end{figure*}

The effects of using various proton momenta in energy dependent cross section  
are shown in Figs.~\ref{edep-full} and \ref{edep-sep} for the 
calculations in the 2-body\_ea scheme with $Q^2=0.06$ (GeV/c)$^2$, 
$\varepsilon = 0.7$, $\Phi_K = 180^\circ$, and 
$\theta_{Ke}=6^\circ$ changing the electron kinematics 
accordingly. The notation is the same as in 
Figs.~\ref{adep-full} and \ref{adep-sep}, respectively.
%
%
%
%

In Figure~\ref{edep-full} we observe significant effects for both 
elementary amplitudes, especially at energies above 2 GeV. 
The resonant nature of the results for the states 
$2^-$, $1^+$, and $3^+$ comes from the longitudinal contributions, 
as shown in Fig.~\ref{edep-sep}, and this structure is strongly 
controlled by the selection rule. A character of the results, 
{\it i.e.} ordering  of the curves, is largely similar above 2 GeV 
and for a given amplitude 
but there are differences in a magnitude of the effects and in a particular 
shape of the curves. The shape depends quite strongly on the elementary 
amplitude but it also differs for the two groups of hypernuclear states 
which is related to the selective contribution from ${\cal F}^1_{00}$.
Larger Fermi effects for SLA are observed in kinematics with 
$Q^2=0.01$ (GeV/c)$^2$ and $E_\gamma=1.5$ GeV than here, which is given 
by a magnitude of the effects in the spin-flip amplitude 
${\cal F}^1_{\lambda\eta}$. In the case of the BS3 amplitude big effects 
are observed for both kinematics at energies above 2 GeV which is given 
by rising strength in the BS3 amplitude at higher energies as one can see  
in Figs. 5 and 11 of Ref.~\cite{SB18}. For example, at $E_\gamma = 2.4$ GeV 
the proton-photon invariant energy is 2.3 and 2.6 GeV in the frozen-proton 
and frozen-$\Lambda$ approximations, respectively, which makes the 
hypernuclear cross sections significantly larger for the latter. As 
the SLA amplitude provides almost a constant strength~\cite{SB18} 
around 2 GeV, the Fermi effects are moderate in this energy region. 
Generally, the results with BS3 reveal more pronounced structures in the 
energy dependence than those with SLA. The structures are quite similar 
within the two groups of states, which is mainly driven by the contribution 
from the spin-flip part ${\cal F}^1_{\lambda\eta}$. 

The resonant nature of the energy-dependent full cross section observed 
mainly for BS3 is induced by contributions from the longitudinal (L) and 
interference (TL) parts as one can clearly see in Fig.~\ref{edep-sep}.  
The transverse part $d\sigma_T$ reveals a smooth energy dependence 
but the contributions from the L and TL parts model the resonant 
behavior in the full cross section, see the black (frozen proton) 
and red (optimum) lines for the $3^+$ state. 
Note that the behavior of the L and TL contributions significantly 
differs for the two extreme cases with the mean momentum, 
$\theta_{p}= 0^\circ$  and  $180^\circ$, which suggests that 
the contributions from the L and TL parts strongly depend on the angle 
between the proton and photon momenta.

Even if the energy behavior of the L and TL parts is quite similar for 
both states, see for example the results with the frozen proton and 
optimum momentum, the magnitude of their contributions differ quite 
strongly. The full cross sections at $E_\gamma = 2$ GeV with frozen proton 
for the $1^-$ and $3^+$ states is about 30 and 150 nb/sr, respectively, 
and the contributions from L are about 1 and 60 nb/sr, respectively. 
The relative contribution from L then amounts to about 3\% and 40\% for 
the $1^-$ and $3^+$ states, respectively. This phenomenon can be 
attributed to a significant contribution from the elementary amplitude 
${\cal F}^1_{00}$ in the longitudinal part for the $3^+$ state (where 
$A^0_{30}\ne 0$) which is absent in $1^-$ (where $A^0_{10}= 0$). Note, 
however, that the contribution from L is smaller for energies different 
from approximately 2~GeV. The relatively large contribution from the 
longitudinal part at 2~GeV demonstrates again importance of the 
electroproduction calculations even in kinematics with a photon which 
is almost real ($Q^2=0.06$ (GeV/c)$^2$). 

In Figures~\ref{adep-full}--\ref{edep-sep} one sees that the results 
with the optimum proton momentum generally lie between the extreme cases 
with frozen proton and frozen $\Lambda$ and in most cases are close to the 
frozen proton approximation. Moreover, because in the version with the 
optimum momentum all the computational schemes are equivalent and calculations 
are performed with only one value of the kaon momentum, we suggest this 
version to be a suitable variant of the optimal factorization 
approximation in the DWIA and denote it as 
{\it optimum on-shell approximation}.  
%
\subsection{Comparison with data and previous results}
In Table~\ref{TabBS3} we compare the new theoretical predictions for the 
triple-differential cross sections in electroproduction of $^{12}_{~\Lambda}$B  
with experimental data and our previous results (``old''), both from Table III 
of Ref.~\cite{archival}. The calculations are performed with the BS3 elementary 
amplitude in the following kinematics:  $E_e=$ 3.77 GeV, $E_e^\prime=$ 1.56 GeV, 
$\theta_e=$ 6$^\circ$, $\theta_{Ke}=$ 6$^\circ$, $\Phi_{K}=$ 180$^\circ$. 
The photon kinematics is $E_\gamma=$ 2.21 GeV, $Q^2=$ 0.0644 (GeV/c)$^2$, 
$\varepsilon=$ 0.7033, the photon angle with respect to the beam is 
4.2$^\circ$, and $\Gamma=$ 0.0178 (GeV/c)$^{-1}$ .  
The kaon angle with respect to the photon momentum is very small 
$\theta_{K}=$ 1.8$^\circ$.

The new results ``NEWa'' are calculated in the frozen-proton approximation and 
the 2-body\_ea scheme with two values of the kaon momentum  
$|\vec{p}_K(2b)|=1931$ MeV/c and $|\vec{p}_K(mb)|=1964$ MeV/c. 
The momentum of $\Lambda$ is 301 MeV/c and it equals the momentum transfer.
The result ``NEWb'' is the same calculation as NEWa but with the optimum proton 
momentum $|\vec{p}_{\sf opt}|=$ 99 MeV/c and cos($\theta_{p\Delta}$)= $-1$. 
The kaon momentum is $|\vec{p}_K(2b)|= |\vec{p}_K(mb)|=$ 1964 MeV/c, 
the $\Lambda$ moves with momentum 170 MeV/c and the momentum transfer is 
$|\vec{\Delta}|=$ 269 MeV/c. One sees that the optimum momentum is comparable 
to the $\Lambda$ momentum and to  $|\vec{\Delta}|$. We denote the calculation 
NEWb as the optimum on-shell approximation, see also Sec.~\ref{optimum}.  
%
%
\begin{table}[hbt]
\begin{tabular}{crcrrrr}
 \hline 
\multicolumn{2}{c}{Data}  & \multicolumn{5}{c}{Theoretical predictions}\\
$E_x$ &  crs\ \ & $E_x$\ \ &\  J$^P$ &  \multicolumn{3}{c}{Cross sections} \\
(MeV) &         & (MeV)    &       &  old\ \    &\  NEWa  &\   NEWb  \\ 
    \hline
      &      & 0.000  & 1$^-$ & 0.524  & 0.611 &  0.741 \\
      &      & 0.116  & 2$^-$ & 2.172  & 2.535 &  2.677 \\
 0.0  & 4.51 & \ \ sum:       &       & 2.696  & 3.145 &  3.418 \\
  \hline
      &      & 2.587  & 1$^-$ & 0.689  & 0.805 &  0.956 \\ 
      &      & 2.593  & 0$^-$ & 0.071  & 0.082 &  0.027 \\
 2.62 & 0.58 & \ \ sum:       &       & 0.760  & 0.887 &  0.983 \\
  \hline 
      &      & 4.761  & 2$^-$ & ---    & 0.022 &  0.022 \\
      &      & 5.642  & 2$^-$ & 0.359  & 0.422 &  0.429 \\ 
      &      & 5.717  & 1$^-$ & 0.097  & 0.113 &  0.132 \\ 
 5.94 & 0.51 & \ \ sum:       &       & 0.456  & 0.558 &  0.583 \\
  \hline 
      &      & 10.480 & 2$^+$ &\ \ 0.157  & 0.175 &  0.196 \\
      &      & 10.525 & 1$^+$ & 0.100  & 0.111 &  0.098 \\
      &      & 11.059 & 2$^+$ & 0.778  & 0.870 &  0.973 \\
      &      & 11.132 & 3$^+$ & 1.324  & 2.169 &  2.099 \\
      &      & 11.674 & 1$^+$ & 0.047  & 0.085 &  0.087 \\
10.93 &\ \  4.68 & \ \ sum:       &       & 2.406  & 3.410 &  3.453 \\
 \hline 
      &      & 12.967 & 2$^+$ &  0.447 & 0.504 &  0.556 \\
      &      & 13.074 & 1$^+$ &  0.196 & 0.219 &  0.191 \\
      &      & 13.383 & 1$^+$ &    --- & 0.0008 &  0.0008\\
12.65 & 0.63 & \ \ sum:       &       &  0.643 & 0.724  &  0.748 \\ 
 \hline 
\end{tabular}
\caption{Experimental cross sections (crs) for electroproduction of 
$^{12}_{~\Lambda}$B in nb/(sr$^2$GeV) are compared with theoretical 
predictions calculaterd with the BS3 amplitude. The data and results 
``old'' are from~\cite{archival}. The newly elaborated results (see 
Sec.~\ref{calculations}) ``NEWa'' 
and ``NEWb'' are calculated with the zero and the optimum proton momentum, 
respectively. The theoretical prediction on the line denoted with ``sum:'' 
is the total cross sections for given multiplet which can be compared with 
the experimental value.}
\label{TabBS3}
\end{table}  

The differences between the results ``old'' and ``NEWa'' observed in 
Table~\ref{TabBS3} are due to our improvements in the model calculations, 
especially in the radial integrals, see details of the calculation in 
Sec.~\ref{calculations}. A remarkable change is the markedly larger cross 
section for the state 3$^+$ at 11.132~MeV which is mainly due to a flaw 
in the previous calculations. The new value significantly improves agreement 
with the data for the second dominant peak at 11~MeV. We can say that 
at the given energy the new cross sections for all states are larger by 10--15 \% 
for both BS3 and SLA elementary amplitudes.

On the other hand, differences between the results with the zero (NEWa) 
and the optimum proton momentum (NEWb) depend quite strongly on the elementary 
amplitude. The new results with the optimum proton momentum are larger by 1--11\% 
for BS3 as shown in Table~\ref{TabBS3} but the results with SLA are larger only 
by less than 1\%. However, recall that the effects from the proton motion for 
SLA are larger for energies different from 2.21~GeV, {\it e.g.} for 
$E_\gamma=$ 1.5 and 2.4 GeV as seen in Fig.~\ref{edep-full}. 
Note also that the Fermi-motion effects in Table~\ref{TabBS3} are moderate 
due to a quite small value of the optimum proton momentum (99 MeV/c). 

We can conclude that both the effects of the proton motion and the improvements 
of our model calculations lead to a better agreement of theoretical predictions 
NEWb with the data for both main peaks at 0 and 11 MeV. The new results are still 
about 27\% below the experimental values for the main peaks whereas the new 
calculation overpredicts the cross sections for the other core-excited states. 
Let us note, however, that in the optimum on-shell approximation one uses only 
one value of the kaon momentum which allows for both the energy conservation in  
the overall system and the use of the on-energy-shell elementary amplitude. 
Recall also that the optimum momentum is not determined uniquely as it also 
depends on the angle between the proton momentum and the momentum transfer 
$\vec{\Delta}$. In our analysis we have considered only the special case with 
$\cos(\theta_{p\Delta})= -1$. 

For completeness, in Tables~\ref{TabBS3-Li} and \ref{TabBS3-N} we also report 
on the new calculations
with the BS3 amplitude for the hypernuclei $^9_\Lambda$Li and $^{16}_{~\Lambda}$N. 
These old results and the experimental data were already 
published in Ref.~\cite{archival} and here we compare them with the new results 
similarly to what we did for $^{12}_{~\Lambda}$B in Table II. One can see that the 
Fermi-motion effects due to optimum proton momentum raise the cross sections 
by 3--20\% for $^{16}_{~\Lambda}$N ($p_{\sf opt}$= 120 MeV/c) and only by 4--6\%  
for $^9_\Lambda$Li ($p_{\sf opt}$= 82 MeV/c). The smaller effects in the latter case
can be partially attributed to the smaller value of the proton momentum. 
These effects together with improvements in the calculation make disagreement 
of the theoretical results with the data still worse for $^{16}_{~\Lambda}$N  
but they improve the agreement of theory and data for $^9_\Lambda$Li.
Note that in the calculations with SLA, changes of the cross sections due
to improvements are of similar magnitude but that the Fermi-motion effects 
are smaller.

%
%
\begin{table}[hbt]
\begin{tabular}{crcrrrr}
 \hline 
\multicolumn{2}{c}{Data}  & \multicolumn{5}{c}{Theoretical predictions}\\
$E_x$ &  crs\ \ & $E_x$\ \ &\  J$^P$ &  \multicolumn{3}{c}{Cross sections} \\
(MeV) &         & (MeV)    &       &  old\ \    &\  NEWa  &\   NEWb  \\ 
    \hline
 0.0  & 0.59 & 0.000  & 3/2$^+$ & 0.157  & 0.188 &  0.197 \\
 0.57 & 0.83 & 0.563  & 5/2$^+$ & 1.035  & 1.238 &  1.314 \\
 sum: & 1.42 &        &         & 1.19   & 1.43  &  1.51  \\
  \hline
      &      & 1.423  & 1/2$^+$ & 0.294  & 0.353 &  0.399 \\ 
      &      & 1.445  & 3/2$^+$ & 0.343  & 0.412 &  0.398 \\
 1.45 & 0.79 & \ \ sum: &       & 0.64   & 0.77  &  0.80  \\
  \hline 
      &      & 2.272  & 5/2$^+$ & 0.109  & 0.130 &  0.147 \\
      &      & 2.732  & 7/2$^+$ & 0.315  & 0.379 &  0.382 \\  
 2.27 & 0.54 & \ \ sum: &       & 0.42   & 0.51  &  0.53  \\
 \hline 
\end{tabular}
\caption{The same as in Table~\ref{TabBS3} but for $^9_\Lambda$Li.} 
\label{TabBS3-Li}
\end{table}  
Different results were obtained for $^{12}_{~\Lambda}$B at lower photon 
energy $E_\gamma = 1.5$ GeV in kinematics of the Hall C experiment~\cite{HallC} 
where the Fermi-motion effects raise the cross sections by 2--4\% for SLA but they 
lower them by about 6\% for BS3. Note that the effects from improvements of 
the calculations are also important at this energy raising the cross sections 
which improves agreement of theoretical predictions with the data. 
This is illustrated in Table~\ref{TabSLA-C} where we compare the new 
results in the frozen-proton (NEWa) and optimum on-shell (NEWb) approximations with 
data from the Hall C experiment E01-011 and with our previous results published 
in Table VI of Ref.~\cite{archival}. We show only the results for $\Lambda$ 
in the $s$ orbit for which assignment of the theory and experimental data is 
straightforward. The differential cross sections were calculated at kinematics, 
$E_e= 1.851$ GeV, $E_e^\prime =0.351$ GeV, $\theta_e = 5.4^\circ$, 
$\theta_{Ke} = 7.11^\circ$ and $\Phi_K= 90^\circ$, with 
the elementary amplitude SLA~\cite{SLA} for which we obtained a better 
agreement with the data than with the BS3 amplitude. The new results with 
the optimum proton momentum (128 MeV/c) are in a better agreement 
with the data than the results with zero proton momentum. However, significant 
improvement comes from the changes in the model calculations, 
compare the results ``old" and ``NEWa". 

%
%
\begin{table}[hbt]
\begin{tabular}{crcrrrr}
 \hline 
\multicolumn{2}{c}{Data}  & \multicolumn{5}{c}{Theoretical predictions}\\
$E_x$ &  crs\ \ & $E_x$\ \ &\  J$^P$ &  \multicolumn{3}{c}{Cross sections} \\
(MeV) &         & (MeV)    &       &  old\ \    &\  NEWa  &\   NEWb  \\ 
    \hline
      &      & 0.000  & 0$^-$ & 0.134  & 0.151 &  0.071 \\
      &      & 0.023  & 1$^-$ & 1.391  & 1.587 &  2.027 \\
  0.0 & 1.45 & \ \ sum: &     & 1.52   & 1.74  &  2.10  \\
  \hline
      &      & 6.730  & 1$^-$ & 0.688  & 0.800 &  0.967 \\ 
      &      & 6.978  & 2$^-$ & 2.153  & 2.502 &  2.679 \\
 6.83 & 3.16 & \ \ sum: &     & 2.84   & 3.30  &  3.65  \\
  \hline 
      &      & 11.000 & 2$^+$ & 1.627  & 1.777 &  2.078 \\
      &      & 11.116 & 1$^+$ & 0.679  & 0.767 &  0.752 \\ 
      &      & 11.249 & 1$^+$ & 0.071  & 0.064 &  0.058 \\ 
10.92 & 2.11 & \ \ sum: &     & 2.38   & 2.61  &  2.89  \\
  \hline
      &      & 17.303 & 1$^+$ & 0.181  & 0.197 &  0.183 \\
      &      & 17.515 & 3$^+$ & 2.045  & 3.597 &  3.550 \\
      &      & 17.567 & 2$^+$ & 1.723  & 1.910 &  2.142 \\
17.10 & 3.44 & \ \ sum: &     & 3.95   & 5.70  &  5.88  \\
 \hline 
\end{tabular}
\caption{The same as in Table~\ref{TabBS3} but for $^{16}_{~\Lambda}$N.} 
\label{TabBS3-N}
\end{table}  
%
%
\begin{table}[hbt]
\begin{tabular}{crcrrrr}
 \hline 
\multicolumn{2}{c}{Data}  & \multicolumn{5}{c}{Theoretical predictions}\\
$E_x$ &  crs\ \ & $E_x$\ \ &\  J$^P$ &  \multicolumn{3}{c}{Cross sections} \\
(MeV) &         & (MeV)    &       &  old\ \    &\  NEWa  &\   NEWb  \\ 
    \hline
      &      & 0.000  & 1$^-$ & 13.90  & 17.89 & 18.04  \\
      &      & 0.116  & 2$^-$ & 44.70  & 57.48 & 60.00  \\
 0.0  & 101.0 & \ \ sum: &    & 58.60  & 75.37 & 78.04  \\
  \hline
      &      & 2.587  & 1$^-$ & 17.26  & 22.20 & 23.05  \\ 
      &      & 2.593  & 0$^-$ &  0.04  &  0.05 &  0.01  \\
 3.127& 33.5 & \ \ sum: &     & 17.31  & 22.25 & 23.06  \\
  \hline 
      &      & 4.761  & 2$^-$ &  0.37  &  0.49  &  0.50 \\
      &      & 5.642  & 2$^-$ &  7.20  &  9.37  &  9.74 \\ 
      &      & 5.717  & 1$^-$ &  2.44  &  3.17  &  3.23 \\ 
 6.077& 26.0 & \ \ sum: &     & 10.01  & 13.03  & 13.47 \\
  \hline
\end{tabular}
\caption{Differential cross sections in nb/sr (crs) from the Hall C 
experiment E01-011 for electroproduction of $^{12}_{~\Lambda}$B are 
compared with theoretical predictions calculated with the SLA 
amplitude. Only the states with $\Lambda$ in the $s$ orbit are shown.   
The data are from \cite{HallC} and the results ``old'' are 
from~\cite{archival}. The new results ``NEWa'' and ``NEWb'' 
are calculated with the zero and optimum proton momentum, respectively. 
The theoretical prediction on the line denoted with ``sum:'' 
is the total cross section for the given multiplet which can be 
compared with the experimental value.}
\label{TabSLA-C}
\end{table}  
%
%
\section{Summary}
We investigated the cross sections for electroproduction of $^{12}_{~\Lambda}$B  
and effects therein, which are caused by various options for kaon and proton 
momenta. 
The value of the kaon momentum is related to a choice of the computational 
scheme and the proton momentum relates to the Fermi motion of the target 
proton inside of the nucleus. The calculations 
were performed in the DWIA where the many-particle matrix element is treated 
in the optimal factorization approximation with an effective proton momentum 
in the elementary amplitude. In order to allow for non-zero values of the proton 
momentum, we have derived the two-component (CGLN-like) form of the elementary 
amplitude which applies generally for electroproduction of pseudo-scalar 
mesons off the nucleons. These CGLN-like amplitudes in a general reference 
frame can be calculated from known scalar amplitudes. Note that in the previous 
calculations the approximation with zero proton momentum (frozen proton) 
was considered and that, as far as we know, this general form of the 
electroproduction amplitude was not available in literature yet.   

Utilizing this new formalism for the elementary amplitude with the 
nucleus-hypernucleus transition matrix elements (OBDME) and the kaon 
distortion as in our previous calculations~\cite{archival} we have 
found that the effects of various choices of the proton effective momentum 
(the Fermi motion effect) depend quite strongly on kinematics and 
the elementary amplitude. 
The effects are in general more pronounced for larger photon energies, 
{\it i.e.} above 2~GeV in the target-nucleus laboratory frame, 
which is given by the magnitude of the effects 
in the elementary amplitude. Generally, larger effects were observed for 
the BS3 amplitude, especially in the energy dependence of the cross sections. 
Resonant structures in the full cross section are modeled by contributions 
from the longitudinal (L) and transverse-longitudinal interference (TL) parts 
which reveal quite a strong sensitivity to the value of proton momentum.  
 
The Fermi motion effects also differ for the hypernuclear states with various  
spins and parities due to the selective contribution from the strong longitudinal 
spherical elementary amplitude ${\cal F}^1_{00}$. We denote these noticeable 
differences of the results for two groups of hypernuclear states as the dynamical 
selection rule which enters into the full cross section mainly via the L and TL 
contributions. We can therefore conclude that the Fermi motion effects are 
very important for the L and TL contributions in the full cross section. 

We have also shown that the cross section depends quite strongly on a scheme 
of computing the kaon laboratory momentum. 
If the kaon momentum is computed from the energy conservation in the 
elementary-production vertex, the elementary amplitude is on-energy-shell 
and the energy conservation of the overall system is violated by about 1\%. 
On the contrary, calculations with the kaon momentum from the overall energy 
conservation have big effects in the elementary amplitude which is 
off-energy shell. 
The off-shell value of the dominant amplitude ${\cal F}^1_{00}$ differs 
significantly from the on-shell value resulting in large contributions 
in the L and TL components of the full cross section.

As the off-energy-shell extension of the elementary amplitude is generally 
not well under control, we prefer using the computational schemes with the 
on-energy-shell elementary amplitude. In the previous calculations we therefore 
considered a hybrid form with the on-energy-shell elementary amplitude and with 
the kaon momentum calculated from the overall energy conservation used 
in the remaining parts of the computation, {\it i.e.} two different kaon 
momenta were used. 
However, the new formalism for the elementary amplitude developed here allows 
to use an ``optimum'' proton momentum which makes the computational schemes 
equivalent, fulfilling both energy conservation relations with one value of 
the kaon momentum, and also allows us to use the on-energy-shell elementary amplitude.

This optimum on-shell approximation was shown to be a suitable 
choice of the effective proton momentum in the optimal factorization 
approximation in DWIA as the obtained results are in a better agreement 
with the experimental data.
Note, however, that this improvement is also partially related to elaborating our 
model calculations. The optimum proton momentum used here for comparison 
with the experimental data on the $^{12}$C target amounts to about 100 MeV/c 
which is in a reasonable agreement with 
the mean value of the proton momentum in the $p$ orbit in $^{12}$C, 
about 180 MeV/c. We therefore suggest using this optimum on-shell 
approximation in DWIA calculations.  

Note that as the optimum momentum is not determined uniquely, one could also 
try other values of the angle of the proton momentum with respect to the 
momentum transfer. The option used here with the proton moving opposite to 
the momentum transfer minimizes the momentum of the $\Lambda$. 
%
%
%
\section*{ACKNOWLEDGMENTS}
The authors thank John Millener for useful discussions and Avraham Gal 
for careful reading of the manuscript. P. B. also acknowledge the warm 
hospitality at INFN Sezione di Roma~1, Gruppo Sanita, where the work 
was initiated. This work was supported by the Czech Science Foundation 
GACR Grant No. 19-19640S.

%
%
%
\appendix
\begin{widetext}
\section{The CGLN-like amplitudes in a general reference frame}
\label{CGNL}
The CGLN-like amplitudes in Eq.~(\ref{general-amplitude}) depend on the six scalar 
amplitudes $A_j$~\cite{SB18} and four-vectors $q=(q_0,\vec{q}\,)$, 
$p_p=(E_p,\vec{p}_p)$, and $p_\Lambda=(E_\Lambda,\vec{p}_\Lambda)$. 
We denote the scalar product as $(a\cdot b)= a_0b_0 -\vec{a}\,\vec{b}$. 
The amplitudes are normalised with  
$N=1/\sqrt{4 m_\Lambda m_p (E_\Lambda +m_\Lambda)(E_p+m_p)}$ where 
$m_p$ and $m_\Lambda$ are proton and $\Lambda$ masses, respectively. 
Except for the photon (in electroproduction) the particles are on the mass 
shell, $m^2= E^2-\vec{p}^{\,2}$. The CGLN-like amplitudes in terms of 
the scalar amplitudes and kinematical variables read

\begin{eqnarray}
G_1 &=& 
  N \left\{\frac{}{}\!\!\left[(p_\Lambda\cdot p_p)\,q_0 -m_\Lambda 
  m_p\,q_0 -(q\cdot p_\Lambda)(E_p+m_p) -(q\cdot p_p)(E_\Lambda 
  +m_\Lambda)\right] A_1 \right. \nonumber\\
 && \left. \ \ \ \ +\left[(p_\Lambda\cdot p_p)+m_\Lambda m_p 
  +E_\Lambda m_p +E_p m_\Lambda \right] \left[(q\cdot p_p)A_4 
  +(q\cdot p_\Lambda)A_5 -q^2 A_6 \right]\frac{}{}\!\!\right\}
  \,, \nonumber\\
G_2 &=& 
  N \left[\,(q_0+E_p+m_p-E_\Lambda-m_\Lambda )A_1 +(q\cdot p_p)A_4 
  +(q\cdot p_\Lambda)A_5 -q^2 A_6\right]\,,  \nonumber\\
G_3 &=& 
  N \left[-(E_p+m_p)A_1\,\right]\,, \nonumber\\
G_4 &=& 
  N \left[-q_0\,A_1 -(q\cdot p_p)A_4 -(q\cdot p_\Lambda)A_5 +q^2 A_6\,
  \right]\,, \nonumber\\
G_5 &=& 
  N \left[-A_5 +A_6\,\right]\,,  \nonumber\\
G_6 &=& 
  N \left[-A_4 -A_5\,\right]\,,  \nonumber\\
G_7 &=& 
  N \left[\,A_5\,\right]\,,  \nonumber\\
G_8 &=& 
  N \left\{\frac{}{}\!\!-(E_p+m_p)A_1 -\frac{E_p+m_p}{q^2} \left[ (q
  \cdot p_p)A_2 + ((q\cdot p_\Lambda)-q^2)A_3\right] \right.\nonumber\\
 && \left. \ \ \ \ +\left[(p_\Lambda\cdot p_p) -(q\cdot p_p) +m_p(E_\Lambda 
  +m_\Lambda -q_0) +E_p\,m_\Lambda \right](A_5 -A_6) \frac{}{}\!\right\} 
   \,, \nonumber\\
     \end{eqnarray} 
 \begin{eqnarray}
G_9 &=& 
  N \left\{\frac{}{}\!(q_0-E_p-m_p-E_\Lambda-m_\Lambda)A_1 +(E_p+m_p)
  (A_2 +A_3) \right.\nonumber\\
 && \left.\ \ \ \ +\left[(p_p\cdot p_\Lambda)+m_p(E_\Lambda +m_\Lambda 
  -q_0) +E_p\,m_\Lambda \right]A_4 \right. \nonumber\\ 
 && \left.\ \ \ \ +\left[(q\cdot p_\Lambda) - (q\cdot p_p) 
  +(p_p\cdot p_\Lambda) +m_p(E_\Lambda +m_\Lambda -q_0) 
  +E_p m_\Lambda \right]A_5 -q^2 A_6 \frac{}{}\!\right\}\,, \nonumber\\
G_{10} &=& 
  N \left\{\frac{}{}\!(E_p+m_p)(A_1 -A_3) + 
  \left[(q\cdot p_p) -(p_p\cdot p_\Lambda)-m_p(E_\Lambda +m_\Lambda -q_0) 
  -E_p\,m_\Lambda \right]A_5 \right\}\,,  \nonumber\\
G_{11} &=& 
   N \left\{ \frac{}{}\!\!q_0\,A_1 +\frac{1}{q^2}(E_\Lambda +m_\Lambda 
  -E_p -m_p) 
  \left[ (q\cdot p_p)A_2 + ((q\cdot p_\Lambda)-q^2)A_3 \right] 
   +(q\cdot p_p)A_4 -\right. \nonumber\\
  && \left.\ \ \ \ -\left[ (q\cdot p_p) +q_0(m_\Lambda +m_p)\right]A_5 
  +\left[ (q\cdot p_p) +(q\cdot p_\Lambda) -q^2 +q_0(m_\Lambda +m_p) 
  \right] A_6 \frac{}{}\!\right\}\,, \nonumber\\   
G_{12} &=& 
   N \left\{ \frac{}{}\!2\,q_0\,A_1 +(E_p +m_p -E_\Lambda -m_\Lambda)
   (A_2+A_3) -q_0(m_\Lambda +m_p)(A_4+A_5) \right.\nonumber\\
   && \left. -\left[(q\cdot p_\Lambda)-(q\cdot p_p)\right] 
   (A_4-A_5) -2\,q^2A_6 \frac{}{}\!\right\}\,,  \nonumber\\
G_{13} &=& 
   N \left\{ \frac{}{}\!\!\!-q_0\,A_1 +(E_\Lambda +m_\Lambda -E_p -m_p)A_3  
   - (q\cdot p_p)A_4 +\left[ (q\cdot p_p) +q_0(m_\Lambda +m_p)\right]A_5 
   + q^2 A_6 \frac{}{}\!\right\}\,, \nonumber\\
G_{14} &=& 
   N \left\{\frac{E_p+m_p}{q^2} \left[ (q\cdot p_p)A_2 + 
  ((q\cdot p_\Lambda)-q^2)A_3\right] +\left[(q\cdot p_p)+q_0m_p\right]
  (A_5-A_6)\frac{}{}\!\right\}\,,  \nonumber\\
G_{15} &=& 
   N \left\{ \frac{}{}\!\!\!-q_0\,A_1-(E_p+m_p)(A_2+A_3) +q_0\,m_p A_4 
  +\left[(q\cdot p_p) +q_0\,m_p -(q\cdot p_\Lambda)\right]A_5 
  +q^2 A_6 \frac{}{}\!\right\}\,,  \nonumber\\
G_{16} &=& 
   N \left\{ \frac{}{}\!\! (E_p +m_p)A_3 - 
    \left[(q\cdot p_p) +q_0\,m_p\right] A_5 \frac{}{}\!\right\}\,. \nonumber
\end{eqnarray}
%
%
%
\section{The spherical amplitudes in a general reference frame}
\label{spherical}
The non spin-flip ($S=0$) spherical amplitudes ${\cal F}^S_{\lambda\eta}$ 
in Eq.~(\ref{spherical-ampl}) can be written in terms of the CGLN-like 
amplitudes and spherical components of the photon ($\vec{q}$), 
proton ($\vec{p}_p$) and kaon ($\vec{p}_K$) momenta
\begin{eqnarray}
{\cal F}^0_{-10}&=& -|\vec{q}\,|\,[\,(p_p)^{\;1}_{-1}\,G_2 
+(p_K)^{\;1}_{-1}\,G_3\,]+
[\,(p_p)^{1}_{0}\;(p_K)^{\;1}_{-1}-(p_p)^{\;1}_{-1}\,(p_K)^{1}_{0}\,]\,
G_4+\nonumber\\
&& +D\,[\,(p_p)^{\;1}_{-1}\,G_6 +(p_K)^{\;1}_{-1}\,G_7\,]\nonumber\\
{\cal F}^0_{\,00}&=& 
-[\,(p_p)^{\,1}_{-1}\;(p_K)^{1}_{1}-(p_p)^{1}_{1}\,(p_K)^{\,1}_{-1}\,]\,
G_4+D\,[\,|\vec{q}\,|\,G_5 + (p_p)^{1}_{0}\,G_6 +(p_K)^{1}_{0}\,G_7\,]
\nonumber\\
{\cal F}^0_{\,10}&=& 
|\vec{q}\,|\,[\,(p_p)^{1}_{1}\,G_2 +(p_K)^{1}_{1}\,G_3\,]-
[\,(p_p)^{1}_{0}\;(p_K)^{1}_{1}-(p_p)^{1}_{1}\,(p_K)^{1}_{0}\,]\,G_4+
\nonumber\\
&& +D\,[\,(p_p)^{1}_{1}\,G_6 +(p_K)^{1}_{1}\,G_7\,]\nonumber
\end{eqnarray}
Similarly we can write down the spin flip ($S=1$) spherical amplitudes
\begin{eqnarray}
{\cal F}^1_{11} &=& 
 G_1 -(p_p)^{\,1}_{-1}\,[\,(p_p)^{1}_{1}\,G_{12}+(p_K)^{1}_{1}\,G_{13}\,] 
  -(p_K)^{\,1}_{-1}\,[\,(p_p)^{1}_{1}\,G_{15}+(p_K)^{1}_{1}\,G_{16}\,]
  \nonumber\\
{\cal F}^1_{10} &=&  
  |\vec{q}\,|\,[\,(p_p)^{1}_{1}\,G_9+(p_K)^{1}_{1}\,G_{10}\,] 
  +(p_p)^{1}_{0}\,[\,(p_p)^{1}_{1}\,G_{12}+(p_K)^{1}_{1}\,G_{13}\,]+
  \nonumber\\
  &&+(p_K)^{1}_{0}\,[\,(p_p)^{1}_{1}\,G_{15}+(p_K)^{1}_{1}\,G_{16}\,]
  \nonumber\\
{\cal F}^1_{1-1} &=& 
  -(p_p)^{1}_{1}\,[\,(p_p)^{1}_{1}\,G_{12}+(p_K)^{1}_{1}\,G_{13}\,] 
  -(p_K)^{1}_{1}\,[\,(p_p)^{1}_{1}\,G_{15}+(p_K)^{1}_{1}\,G_{16}\,]
  \nonumber\\
{\cal F}^1_{01} &=& 
  -|\vec{q}\,|\,[\,(p_p)^{\,1}_{-1}\,G_{11}+(p_K)^{\,1}_{-1}\,G_{14}\,] 
  -(p_p)^{1}_{0}\,[\,(p_p)^{\,1}_{-1}\,G_{12}+(p_K)^{\,1}_{-1}\,G_{15}\,]-
  \nonumber\\
  &&-(p_K)^{1}_{0}\,[\,(p_p)^{\,1}_{-1}\,G_{13}+(p_K)^{\,1}_{-1}\,G_{16}\,]
  \nonumber\\
{\cal F}^1_{00} &=& 
   G_1 + |\vec{q}\,|\,[\,|\vec{q}\,|\,G_8+(p_p)^{1}_{0}\,G_{9} 
   +(p_K)^{1}_{0}\,G_{10} +(p_p)^{1}_{0}\,G_{11}+(p_K)^{1}_{0}\,G_{14}\,]+
   \nonumber\\
  &&+(p_p)^{1}_{0}\,(p_p)^{1}_{0}\,G_{12}+(p_p)^{1}_{0}\,(p_K)^{1}_{0}\,(G_{13}
   +G_{15}) + (p_K)^{1}_{0}\,(p_K)^{1}_{0}\,G_{16}\nonumber\\
{\cal F}^1_{0-1} &=& 
   -|\vec{q}\,|\,[\,(p_p)^{1}_{1}\,G_{11}+(p_K)^{1}_{1}\,G_{14}\,] 
   -(p_p)^{1}_{1}\,[\,(p_p)^{1}_{0}\,G_{12}+(p_K)^{1}_{0}\,G_{13}\,]-
   \nonumber\\
   &&-(p_K)^{1}_{1}\,[\,(p_p)^{1}_{0}\,G_{15}+(p_K)^{1}_{0}\,G_{16}\,]
   \nonumber\\
{\cal F}^1_{-11} &=& 
   -[(p_p)^{\,1}_{-1}\,(p_p)^{\,1}_{-1}\,G_{12} 
   +(p_p)^{\,1}_{-1}\,(p_K)^{\,1}_{-1}\,(G_{13} 
   +G_{15})+(p_K)^{\,1}_{-1}\,(p_K)^{\,1}_{-1}\,G_{16}]\nonumber\\
{\cal F}^1_{-10} &=& 
   |\vec{q}\,|\,[\,(p_p)^{\,1}_{-1}\,G_9+ (p_K)^{\,1}_{-1}\,G_{10}\,] 
   +(p_p)^{1}_{0}\,[\,(p_p)^{\,1}_{-1}\,G_{12}+(p_K)^{\,1}_{-1}\,G_{13}\,]+
   \nonumber\\
  &&+(p_K)^{1}_{0}\,[\,(p_p)^{\,1}_{-1}\,G_{15}+(p_K)^{\,1}_{-1}\,G_{16}\,]
  \nonumber\\
{\cal F}^1_{-1-1} &=& 
   G_1 -(p_p)^{1}_{1}\,[\,(p_p)^{\,1}_{-1}\,G_{12}+(p_K)^{\,1}_{-1}\,
   G_{13}\,] -(p_K)^{1}_{1}\,[\,(p_p)^{\,1}_{-1}\,G_{15} 
   +(p_K)^{\,1}_{-1}\,G_{16}\,]\nonumber
\end{eqnarray}
The spherical components of the momenta and the parameter $D$ are  
\begin{eqnarray}
&&(p_K)^1_0= |\vec{p}_K| \cos\theta_K,\ \ \ \ 
(p_K)^{\;1}_{\pm 1}= \mp \frac{|\vec{p}_K|}{\sqrt{2}}\,\sin\theta_K\,
\exp(\pm i\Phi_K) \ \ \ \ \nonumber\\
&&(p_p)^1_0= |\vec{p}_p| \cos\theta_p,\ \ \ \ 
(p_p)^{\;1}_{\pm 1}= \mp \frac{|\vec{p}_p|}{\sqrt{2}}\,\sin\theta_p\,
\exp(\pm i\Phi_p) \nonumber\\
&&D= \,i\,|\vec{q}\,|\,|\vec{p}_p|\,|\vec{p}_K|\sin\theta_p \sin\theta_K 
(\cos\Phi_p\sin\Phi_K - \sin\Phi_p\cos\Phi_K)\,.\nonumber      
\end{eqnarray}
The polar angles $\theta_K$ and $\theta_p$ are determined with respect 
to the photon momentum and the azimuthal angles $\Phi_K$ and $\Phi_p$ 
are defined with respect to the leptonic plane as shown in Fig.~\ref{frame}. 
Formulas for the CGLN-like amplitudes $G_j$ are given in 
Appendix~\ref{CGNL}.

%
%
\section{Equation for the reduced amplitudes}
\label{details}
Here we briefly show how the equation for the reduced amplitude 
(\ref{amplitude-3}) was obtained. 
We start with the expression (\ref{matrix-5}) for the many-particle matrix 
element in the optimal factorization approximation and consider 
the partial-wave decomposition of the plane waves and the kaon distorted wave 
\begin{equation}
e^{(iB\,\vec{\Delta}\cdot\vec{\xi})}\,\chi^*_K(\vec{p}_{KH},B\vec{\xi}) = 
\sum_{LM} F_{LM}(\Delta B\xi)\,{\sf Y}_{LM}(\hat{\xi})\,,
\end{equation}
where $\xi = |\vec{\xi}|$ is the relative particle-core coordinate and 
$\Delta = |\vec{q} -\vec{p}_K|$ is the momentum transfer. 
Note, that $\vec{p}_{KH}$ is the kaon momentum with respect to the hypernucleus 
and that the radial part $F_{LM}$ also depends on orientation of the momentum 
transfer given by the projection $M$. 
Using this decomposition and the spherical form of the elementary amplitude 
(\ref{spherical-ampl}) in the gauge used in Eq.(\ref{reduction})  
we obtain for the spherical components of the hypernuclear  
production amplitude
\begin{eqnarray}
T_\lambda^{(1)} &=& 
{\sf Z}\,\sum_{LM}\sum_{S\eta}\sum_{Jm} {\sf C}^{Jm}_{LMS\eta}\, 
{\cal F}^S_{\lambda\eta}\!\int\!d^3\xi\,d^3\xi_1..\, d^3\,\xi_{A-2}\;
\Phi^*_H(\vec{\xi}_1,..\,\vec{\xi}_{A-2},\vec{\xi}\,)\,F_{LM}(\Delta B\xi)
\left[Y_L(\hat{\xi})\otimes\sigma^S\right]^J_m\!
\Phi_A(\vec{\xi}_1,...\,\vec{\xi}_{A-2},\vec{\xi}\,) 
\nonumber\\
&=& {\sf Z}\,\sum_{LM}\sum_{S\eta}\sum_{Jm}
{\sf C}^{Jm}_{LMS\eta}\,{\cal F}^S_{\lambda\eta}\,\langle\,\Phi_H\,|\,F_{LM} 
\left[Y_L\otimes\sigma^S\right]^J_m\,|\,\Phi_A\,\rangle\,, 
\end{eqnarray}
where the one-particle transition operator is written as the tensor 
product
 $$F_{LM}\,Y_{LM}\,\sigma^S_\eta = \sum_{Jm}{\sf C}^{Jm}_{LMS\eta} F_{LM} \left[Y_L\otimes\sigma^S\right]^J_m\,,$$
 with ${\sf C}^{Jm}_{LMS\eta}$ the Clebsch-Gordan coefficient.
The nuclear and hypernuclear states are determined by their spin 
($J_A, M_A$) and ($J_H, M_H$), respectively. Utilizing the Wigner-Eckart 
theorem the amplitude is written in terms of reduced matrix elements
\begin{equation}
T_\lambda^{(1)} = 
\frac{{\sf Z}}{\left[J_H\right]}\,\sum_{Jm} {\sf C}^{J_HM_H}_{J_AM_AJm}\,
\sum_{LM}\sum_{S\eta} {\sf C}^{Jm}_{LMS\eta}\, 
{\cal F}^S_{\lambda\eta}\,(\,\Phi_H\,||\,F_{LM} 
\left[Y_L\otimes\sigma^S\right]^J||\,\Phi_A\,)\,.
\label{amplitude-2} 
\end{equation}
   
The reduced matrix element is calculated introducing one-particle states 
$|\alpha\rangle$ with the quantum numbers $|nlj\rangle$ generated by 
the creation operators $|\alpha\rangle=a^+_\alpha|0\rangle$ for the proton 
and $|\alpha^\prime\rangle=b^+_{\alpha^\prime}|0\rangle$ for the $\Lambda$.
Assuming the completeness of the one-particle states and the Wigner-Eckart 
theorem we can write the one-particle operator in this base 
\begin{equation}
F_{LM}\,[Y_L\otimes\sigma^S]^J_m = 
 \frac{1}{{\sf Z}\,[J]}\,\sum_{\alpha\alpha^\prime}\;
(\,\alpha'\,||\,F_{LM}\,[Y_L\otimes\sigma^S]^J\,||\,\alpha\,) 
\;[b_{\alpha'}^+\otimes a_\alpha]^J_m\,,
\label{one-particle-op}
\end{equation}
with the normalisation $\sum b_\alpha^+\,a_\alpha={\sf Z}$ as only 
the protons can be changed to $\Lambda$.  
This form allows us to decompose the many-particle reduced matrix element 
in Eq.~(\ref{amplitude-2}) into the one-particle states 
\begin{equation}
(\,\Phi_H\,||\,F_{LM} 
\left[Y_L\otimes\sigma^S\right]^J||\,\Phi_A\,) = 
\frac{1}{{\sf Z}\,[J]}\sum_{\alpha\alpha^\prime}\;(\,\alpha'\,||\,F_{LM} 
\left[Y_L\otimes\sigma^S\right]^J||\,\alpha\,)
(\,\Phi_H\,||\,[b_{\alpha'}^+\otimes a_\alpha]^J\,||\;\Phi_A\,)\,.
\label{reduced-many-particle}
\end{equation}
The last term in this expression is the reduced one-body density 
matrix element (OBDME) which can be calculated using a nuclear model, 
{\it e.g.} the shell model. In our calculations, OBDMEs and 
the spherical elementary amplitudes ${\cal F}^S_{\lambda\eta}$ 
are the input. 

To calculate the reduced matrix element of the one-particle operator in 
Eq.~(\ref{reduced-many-particle}) we use the one-particle wave functions 
in the coordinate space
\begin{equation}
\langle\,\vec{\xi}\;|\,\alpha\,\rangle = 
\langle\,\vec{\xi}\;|\,n\,l\,j\,\mu\,\rangle =
R_{\alpha}(\xi)\,\sum_{\nu\eta}\,{\sf C}^{j\mu}_{l\nu\frac{1}{2}\eta}\,
Y_{l\nu}(\hat{\xi})\,X_\eta^{\frac{1}{2}}\,,
\end{equation}
where $\vec{\xi}$ is a relative particle-core (Jacobi) coordinate and 
$X_\eta^{\frac{1}{2}}$ is the Pauli spinor. 
 After some manipulations we get
\begin{eqnarray}
(\,\alpha'\,||\,F_{LM} 
\left[Y_L\otimes\sigma^S\right]^J||\,\alpha\,) &=& 
  \frac{1}{\sqrt{2\pi}} [L][S][J][l][l'][j][j']\,
  \left( \begin{array}{ccc}\!l'\ \ L \ \ l\!\\ 0\ \ 0\ \ 0\!
  \end{array} \right) \left\{ \begin{array}{ccc}\!\frac{1}{2}\ \ 
  \frac{1}{2}\ \ S\!\\ j'\ \ j\ \ J\!\\ l'\ \ l\ \ L\!
  \end{array} \right\}\,(-1)^{-l'}\,\nonumber\\
&& \times\,\int_0^\infty \!\!d\xi\,\xi^2\,{R}_{\alpha^\prime}(\xi)^*\,
  F_{LM}(\Delta B\xi)\,{R}_{\alpha}(\xi) \equiv 
  {\cal H}^{LSJ}_{l'j'lj}\;{\cal R}^{LM}_{n'l'nl}\,,
\label{HO-matrix-element}
\end{eqnarray}
where ${\cal H}^{LSJ}_{l'j'lj}$ includes the Racah algebra in 
Eq.~(\ref{HO-matrix-element}) and ${\cal R}^{LM}_{\alpha^\prime\alpha}$ 
is the radial integral which includes the radial parts of 
the one-particle wave functions and the radial part of the transition 
operator. The radial wave functions can be calculated from the 
Schr\"odinger equation with a Woods-Saxon potential as in Ref.~\cite{archival} 
or can be taken consistently from the many-particle calculations of OBDME. 

Combining Eqs. (\ref{amplitude-2}), (\ref{reduced-many-particle}), 
and (\ref{HO-matrix-element}) we obtain the equation   
\begin{equation}
T_\lambda^{(1)} = 
\frac{1}{\left[J_H\right]}\,\sum_{Jm} {\sf C}^{J_HM_H}_{J_AM_AJm}\,
\frac{1}{\left[J\right]}\,\sum_{S\eta}{\cal F}^S_{\lambda\eta}
\sum_{LM} {\sf C}^{Jm}_{LMS\eta}\, 
\sum_{\alpha^\prime\alpha} 
{\cal R}^{LM}_{\alpha^\prime\alpha}\;{\cal H}^{LSJ}_{l'j'lj}\; 
(\,\Phi_H\,||\,[b_{\alpha'}^+\otimes a_\alpha]^J\,||\;\Phi_A\,)\,,
\label{final-ampl} 
\end{equation}
which corresponds to Eqs.~(\ref{amplitude}) and (\ref{amplitude-3}). 
Note that summing over the one-particle states $\alpha^\prime\alpha$ determines 
a model space of the calculation on which the proton $\rightarrow\Lambda$ 
transitions in the OBDMEs are assumed. In the calculations for the $^{12}$C 
target presented here we assume the proton being in the $p$ orbit and 
the $\Lambda$ in the $s$ or $p$ orbit, similarly to our previous calculations in Ref.~\cite{archival}.
\end{widetext}
%
%

\end{document}